\documentclass[reqno,a4paper,11pt]{article}
\pdfoutput=1

\usepackage{xcolor}
\usepackage{graphicx}
\usepackage[textwidth = 430 pt, textheight = 630 pt]{geometry}

\definecolor{MyDarkBlue}{rgb}{0.15,0.25,0.45}
\usepackage{epsfig,rotating}
\usepackage{amsmath,amssymb}
\usepackage{amsfonts}
\usepackage{mathrsfs}
\newcommand\mathbbm\mathbb

\usepackage[normalem]{ulem}

\usepackage{latexsym}
\usepackage{amsthm}
\usepackage[all,knot]{xy}
\xyoption{arc}

\usepackage[utf8x]{inputenc}

\usepackage{hyperref}
\hypersetup{
    hypertexnames=false,
    colorlinks=true,
    citecolor=MyDarkBlue,
    linkcolor=MyDarkBlue,
    urlcolor=MyDarkBlue,
    pdfauthor={Dominik Rist, Christian Saemann and Miro van der Worp},
    pdftitle={Towards an M5-Brane Model III: Self-Duality from Additional Trivial Fields},
    pdfsubject={hep-th math-ph},
    breaklinks=true
}

\usepackage{booktabs}

\usepackage{tikz}
\usetikzlibrary{matrix,cd,arrows}
\usepackage{mathtools}
\usepackage[all,knot]{xy}
\xyoption{arc}

\let\fn\footnote
\renewcommand{\footnote}[1]{\linespread{1.1}\fn{#1}\linespread{1.29}}

\makeatletter\renewcommand{\section}{\@startsection
    {section}{1}{\z@}{-3.5ex plus -1ex minus
        -.2ex}{2.3ex plus .2ex}{\bf }}
\makeatletter\renewcommand{\subsection}{\@startsection{subsection}{2}{\z@}{-3.25ex
        plus -1ex minus
        -.2ex}{1.5ex plus .2ex}{\bf }}
\makeatletter\renewcommand{\subsubsection}{\@startsection{subsubsection}{3}{-2.45ex}{-3.25ex
        plus -1ex minus -.2ex}{1.5ex plus .2ex}{\it }}
\renewcommand{\thesection}{\arabic{section}}
\renewcommand{\thesubsection}{\arabic{section}.\arabic{subsection}}
\renewcommand{\@seccntformat}[1]{\@nameuse{the#1}.~~}

\renewcommand{\theequation}{\thesection.\arabic{equation}}
\makeatletter \@addtoreset{equation}{section}
\def\Ddots{\mathinner{\mkern1mu\raise\p@
        \vbox{\kern7\p@\hbox{.}}\mkern2mu
        \raise4\p@\hbox{.}\mkern2mu\raise7\p@\hbox{.}\mkern1mu}}
\DeclareRobustCommand{\cev}[1]{%
    {\mathpalette\do@cev{#1}}%
}
\newcommand{\do@cev}[2]{%
    \vbox{\offinterlineskip
        \sbox\z@{$\m@th#1 x$}%
        \ialign{##\cr
            \hidewidth\reflectbox{$\m@th#1\vec{}\mkern4mu$}\hidewidth\cr
            \noalign{\kern-\ht\z@}
            $\m@th#1#2$\cr
        }%
    }%
}
\providecommand*{\shuffle}{%
    \mathbin{\mathpalette\shuffle@{}}%
}
\newcommand*{\shuffle@}[2]{%
    \sbox0{$#1\vcenter{}$}%
    \kern .15\ht0 
    \rlap{\vrule height .25\ht0 depth 0pt width 2.5\ht0}%
    \raise.1\ht0\hbox to 2.5\ht0{%
        \vrule height 1.75\ht0 depth -.1\ht0 width .17\ht0 %
        \hfill
        \vrule height 1.75\ht0 depth -.1\ht0 width .17\ht0 %
        \hfill
        \vrule height 1.75\ht0 depth -.1\ht0 width .17\ht0 %
    }%
    \kern .15\ht0 
}

\renewcommand{\thethm}{\thesection.\arabic{thm}}

\newcommand{\appendices}{
    \section*{Appendix}\label{appendices}\setcounter{subsection}{0}
    \addcontentsline{toc}{section}{Appendix}
    \setcounter{equation}{0}
    \makeatletter
    \renewcommand{\theequation}{\Alph{subsection}.\arabic{equation}}
    \renewcommand{\thesubsection}{\Alph{subsection}}
    \renewcommand{\thethm}{\Alph{subsection}.\arabic{thm}}
    \@addtoreset{equation}{subsection}
    \@addtoreset{thm}{subsection}
    \makeatother
}

\makeatother

\newcommand{\makecommand}[3]{%
    \foreach \i in #3 {%
        \expandafter\xdef\csname #1\i\endcsname{\noexpand#2{{\unexpanded\expandafter{\i}}}}%
    }%
}

\newcommand{\latinalphabet}{A,a,B,b,C,c,d,D,E,e,F,f,G,g,H,h,I,i,J,j,K,k,L,l,M,m,N,n,O,o,P,p,Q,q,R,r,S,s,T,t,U,u,V,v,W,w,X,x,Y,y,Z,z}
\newcommand{\upperlatinalphabet}{A,B,C,D,E,F,G,H,I,J,K,L,M,N,O,P,Q,R,S,T,U,V,X,Y,Z}
\makecommand{I}{\mathbbm}{\latinalphabet}
\makecommand{bf}{\mathbf}{\latinalphabet}
\makecommand{bm}{\bm}{\latinalphabet}
\makecommand{ca}{\mathcal}{\upperlatinalphabet}
\makecommand{fr}{\mathfrak}{\latinalphabet}
\makecommand{rm}{\mathrm}{\latinalphabet}
\makecommand{sf}{\mathsf}{\latinalphabet}
\makecommand{sc}{\mathscr}{\latinalphabet}

\makecommand{fr}{\mathfrak}{{der,gl,sl,so,osp,su,sp,spin,string,Poisson}}
\makecommand{sf}{\mathsf}{{id,String,Lie,Hom,SO,SU,Sp,Diff,HSymp,jac,alt,sym,Sym,CE,W,T,TE,End,ChEL,ZE,MC,pr}}

\makecommand{rm}{\mathrm}{{Ham,sk,lp,wk}}

\newcommand{\eps}{\varepsilon}
\newcommand{\ewith}{~~~\mbox{with}~~~}
\newcommand{\eand}{~~~\mbox{and}~~~}

\newcommand{\fsl}[1]{{\ooalign{\(#1\)\cr\hidewidth\(/\)\hidewidth\cr}}}
\def\slasha#1{\setbox0=\hbox{$#1$}#1\hskip-\wd0\hbox to\wd0{\hss\sl/\/\hss}}
\def\slashb#1{\setbox0=\hbox{$#1$}#1\hskip-\wd0\hbox to\wd0{\hss\sl/\/\hss}}

\def\periodb#1{\setbox0=\hbox{$#1$}#1\hskip-\wd0\hbox to\wd0{-}}
\newcommand{\nablas}{\fsl{\nabla}}
\newcommand{\dpar}{\partial}
\newcommand{\dpars}{\slasha{\dpar}}

\newcommand{\comment}[1]{}     				
\def\tyng(#1){\hbox{\tiny$\yng(#1)$}}			
\def\tyoung(#1){\hbox{\tiny$\young(#1)$}}			

\newcommand{\beq}{\begin{eqnarray}}
    \newcommand{\eeq}{\end{eqnarray}}

\let\oldbigotimes\bigotimes
\renewcommand\bigotimes{\oldbigotimes\nolimits}

\newcommand{\aghsk}{\hat{\mathfrak{g}}_{\rm sk}}
\newcommand{\langlec}{\prec\hspace{-0.5mm}}
\newcommand{\ranglec}{\hspace{-0.5mm}\succ}
\newcommand{\epsb}{{\bar{\varepsilon}}}
\newcommand{\chib}{{\bar{\chi}}}
\newcommand{\lambdab}{{\bar{\lambda}}}

\begin{document}
    \begin{titlepage}
        \begin{flushright}
            EMPG--20--25
        \end{flushright}
        \vskip2.0cm
        \begin{center}
            {\LARGE \bf Towards an M5-Brane Model III:\\[0.3cm] Self-Duality from Additional Trivial Fields}
            \vskip1.5cm
            {\Large Dominik Rist, Christian Saemann and Mir\'o van der Worp}
            \setcounter{footnote}{0}
            \renewcommand{\thefootnote}{\arabic{thefootnote}}
            \vskip1cm
            {\em Maxwell Institute for Mathematical Sciences\\
                Department of Mathematics, Heriot--Watt University\\
                Colin Maclaurin Building, Riccarton, Edinburgh EH14 4AS, U.K.}\\[0.5cm]
            Email: \href{mailto:dr40@hw.ac.uk}{\ttfamily dr40@hw.ac.uk}~,~\href{mailto:c.saemann@hw.ac.uk}{\ttfamily c.saemann@hw.ac.uk}~,~\href{mailto:mdv1@hw.ac.uk}{\ttfamily mdv1@hw.ac.uk}
        \end{center}
        \vskip1.0cm
        \begin{center}
            {\bf Abstract}
        \end{center}
        \begin{quote}
            We present a six-dimensional $\caN=(1,0)$ supersymmetric higher gauge theory in which self-duality is consistently implemented by physically trivial additional fields. The action contains both $\caN=(1,0)$ tensor and vector multiplets and is non-trivially interacting. The tensor multiplet part is loosely related to a recently proposed action by Sen that leads to on-shell self-duality in an elegant way. As we also show, Sen's action finds a very natural and direct interpretation from a homotopy algebraic perspective.
        \end{quote}
    \end{titlepage}
    
    \tableofcontents
    
    \section{Introduction and results}
    
    \subsection{Motivation}
    
    This paper is the third in a series of papers~\cite{Saemann:2017zpd,Saemann:2019dsl} attempting to systematically overcome the obstacles for constructing a classical M5-brane model, see also~\cite{Saemann:2017rjm}. There are good reasons to be skeptical about the existence of such a model, cf.~\cite{Lambert:2019khh,Saemann:2017zpd} and references therein. We note, however, that even if this program should fail, many of its results should be applicable and useful in different contexts, such as $F$-theory, heterotic supergravity, double/exceptional field theories, and tensor hierarchies on the physical side as well as higher algebra and differential geometry on the mathematical side.
    
    There are two prominent obstacles towards a construction of an M5-brane model: the correct definition of a higher dimensional parallel transport and implementation of self-duality by an action principle. Both problems have found solutions in the literature but the commonly known solutions are partially unsatisfactory. 
    
    A consistent higher dimensional parallel transport can be defined in the context of higher non-abelian principal bundles or gerbes~\cite{Breen:math0106083,Aschieri:2003mw}, see e.g.~also~\cite{Schreiber:2008aa}. A problem arising here is that consistency of this parallel transport requires a part of the curvature to vanish, which locally renders the connection gauge equivalent to the connection on an abelian gerbe~\cite{Gastel:2018joi,Saemann:2019dsl}. One can circumvent this issue by using special higher gauge algebras which allow for an adjustment, i.e.~essentially an alternative definition of the curvature expressions~\cite{Saemann:2019dsl}. An important example that we will use in our model in this paper are adjusted string structures~\cite{Killingback:1986rd,Redden:2006aa,Sati:2008eg,Waldorf:2009uf}, or rather their metric extension~\cite{Saemann:2019dsl}, which are higher analogues of spin bundles. String structures allow for the definition of an adjusted higher parallel transport~\cite{Kim:2019owc}.
    
    As is well-known by now, self-duality can be implemented by the Pasti--Sorokin--Tonin (PST) mechanism~\cite{Pasti:1995ii,Pasti:1995tn,Pasti:1997gx}. The auxiliary data, however, is somewhat unsatisfactory, particularly when it comes to quantizing such a theory. Moreover, if the PST mechanism is implemented in the models that we are interested in, the Lagrangian comes with terms proportional to inverses of a scalar field~\cite{Bandos:2013jva,Saemann:2017zpd}. We are particularly interested in the limit in which the expectation value of the scalar field vanishes. One can certainly carefully take the relevant limit, but the situation is not ideal. 
    
    An alternative to the PST mechanism was suggested recently by Ashoke Sen~\cite{Sen:2015nph,Sen:2019qit}, see also~\cite{Sen:2015uaa}. It is the primary goal of this paper to incorporate this mechanism into the model considered in~\cite{Saemann:2017zpd} and to extend it suitably to implement all duality relations in a fully supersymmetric fashion.
    
    In a recent paper~\cite{Lambert:2019diy}, Sen's mechanism was used in a Lagrangian whose equations of motion are those of~\cite{Lambert:2010wm} for a non-abelian (2,0)-tensor multiplet. There were attempts at interpreting these equations as higher gauge theories~\cite{Palmer:2012ya,Papageorgakis:2011xg}, but because there are no gauge potential forms in the theory, it remains unclear if a consistent higher parallel transport can be formulated with them.
    
    The model we use in this paper is a six-dimensional $\caN=(1,0)$ supersymmetric field theory~\cite{Saemann:2017zpd}  which is a specialization of the model constructed in~\cite{Samtleben:2011fj,Samtleben:2012fb}. In the latter work, the field theory was constructed using the tensor hierarchy of gauged supergravity. Structure constants and their algebraic constraints were derived from supersymmetry, which also led to equations of motion and ultimately an action principle. The mathematical interpretation of this structure was unclear; it is actually an $EL_\infty$-algebra~\cite{Borsten:2021ljb}. Using the adjusted metric string structures of~\cite{Saemann:2019dsl,Saemann:2017rjm} makes it evident that this theory is mathematically well-defined and consistent: there are obvious notions of connections on higher principal bundles and of a higher parallel transport that underlie the theory.
    
    We hasten to add that this model is clearly not a general M5-brane model: most evidently, it contains a six-dimensional vector multiplet carrying degrees of freedom which cannot be present in an $\caN=(2,0)$ supersymmetric M5-brane model.
    
    It is somewhat mysterious that Sen's comparatively simple mechanism to incorporate self-duality by an action principle has been overlooked for quite some time.\footnote{This may be due to it being seemingly close to Lagrange multipliers, which by themselves are not suitable for implementing self-duality at the quantum level.} A secondary goal of this paper is therefore to clarify the mathematical origin of this mechanism and we use the homotopy algebraic perspective on classical field theories for this, cf.~\cite{Jurco:2018sby}.
    
    \subsection{Results}
    
    We begin in section~\ref{sec:self-duality_and_homotopy_algebras} with an analysis of Sen's mechanism. Recall that any polynomial field equations can be encoded in the homotopy Maurer--Cartan equation of an $L_\infty$-algebra and any polynomial action can be captured by the homotopy Maurer--Cartan action of a corresponding cyclic $L_\infty$-algebra. We find that the $L_\infty$-algebra corresponding to the self-duality equation in six dimensions has a trivial cyclic extension by doubling the field content. This doubling is nothing but the addition of antifields familiar from the BV-formalism. After a field redefinition, i.e.~a strict isomorphism on the cyclic $L_\infty$-algebra, the homotopy Maurer--Cartan action for this theory is precisely Sen's action. In the original papers~\cite{Sen:2015nph,Sen:2019qit}, some work was required to show that parts of the field content decouple and the relevant field content is merely that of a self-dual curvature. In our picture, the cyclic $L_\infty$-algebra allows for a further extension by enhancing the gauge symmetry, which renders the decoupling field gauge trivial. This persists when the action is coupled to additional matter fields in an arbitrary way; in the original papers, only couplings at least cubic in the fields were considered.

    In section~\ref{sec:old_field_theory}, we review in detail the model of~\cite{Saemann:2017rjm,Saemann:2017zpd}. We also comment on an important point which was mostly ignored in the previous work: the particular structure of the $EL_\infty$-algebra underlying the model allows for a cyclic structure which does {\em not} originate from a symplectic form on the grade-shifted graded vector space of the gauge algebra. The latter is usually the case for $L_\infty$-algebras. We also demonstrate the explicit relation and equivalence between the adjusted metric string structures constructed in~\cite{Saemann:2019dsl} and those used in tensor hierarchies and in particular in~\cite{Samtleben:2011fj}.
    
    Our central result is certainly the $\caN=(1,0)$ supersymmetric action, which implements self-duality as well as duality between $2$- and $4$-form curvatures. Its field content is given by a $(1,0)$-tensor multiplet $(B,\chi,\phi)$, a $(1,0)$-vector multiplet $(A,\lambda,Y)$, additional $3$- and $4$-form potentials $C$ and $D$ as well as two auxiliary fields: a self-dual 3-form $\beth_s=\beth_s^+$ and a 2-form $\daleth_t$, see table~\ref{fig:supermultiplets} for more details; further hypermultiplets are readily incorporated, too. The Lagrangian of our action reads as
    \begin{equation}
        \begin{aligned}
            \caL^{\caN=(1,0)}_{\rm TF}&=-\rmd \phi_s\wedge\star \rmd \phi_r -\star4\bar\chi_s\dpars\chi_r+\star4\chib_s\,(\fsl{F}_{\!t},\lambda_t) -\star8\chib^i_s\,(Y_{tij},\lambda^j_t)
            \\  
            &\hspace{1cm}
            +\phi_s\Big((F_t,\star F_t)-\star 2(Y_{tij},Y_t^{ij})+\star4(\bar\lambda_t,\nablas \lambda_t)\Big)
            +\caL_{0,{\rm hyper}}
            \\
            &\hspace{1cm}+(\rmd B_s)^{+} \wedge \left(\rmd B_r + {\rm cs}(A)_r + (\bar{\lambda}_t, \gamma_{(3)} \lambda_t)\right) + (\rmd B_s)^- \wedge C_q^{+}   
            \\
            &\hspace{1cm}-\beth_s \wedge (\rmd B_r + {\rm cs}(A)_r - C_q + (\bar{\lambda}_t, \gamma_{(3)} \lambda_t))
            \\
            &\hspace{1cm}
            +(\nabla C_u)(\daleth_t) +  2B_s \wedge (F_t, \daleth_t) + D_v(\daleth_t) - 2 \phi_s (\daleth_t , \star F_t) 
            \\
            &\hspace{1cm}-\star4\chib_s(\fsl{\daleth}_t,\lambda_t) -B_s \wedge (\daleth_t, \daleth_t) +\phi_s (\daleth_t, \star \daleth_t)~,
        \end{aligned}
    \end{equation}
    where $F_t\coloneqq\rmd A_t+\tfrac12[A_t,A_t]$ and ${\rm cs}(A)_r$ is the Chern--Simons 3-form of $A_t$. Fields with a subscript $t$ and $u$ or $v$ live in a quadratic Lie algebra $\frg$ with bilinear form $(-,-)$ and its dual $\frg^*$, respectively, while fields with a subscript $ r $ or $ q $ (resp. $ s $) live in $ \IR $ (resp. its dual). These subscripts merely help to identify different copies of the same vector space. Indices $ i,j=1,2 $ are standard R-symmetry indices, as explained in appendix \ref{app:susy}, and Einstein summation convention is implied for these indices.  
    
    This action is fully supersymmetric under the transformations given in table~\ref{fig:supermultiplets}, shifted and complemented by the expressions in~\eqref{eq:shifted_SUSY}. We stress that even though the action has some superficial similarities to that of Sen's mechanism~\cite{Sen:2015nph,Sen:2019qit}, it is merely the covariantization and supersymmetrization of a Lagrange multiplier action. Its form is essentially uniquely fixed by supersymmetry. Contrary to Sen's mechanism, we require neither decoupling arguments nor enhanced gauge symmetry to remove the additional degrees introduced by $\beth_s$ and $\daleth_t$. Both fields simply vanish on-shell, cf.~the list of equations of motion in table~\ref{fig:eom}. Note that the additional fields $\beth_s$ and $\daleth_t$ appear in interaction terms, but contrary to Sen's action, there are also interaction terms $(\rmd B_s)^+\wedge\big({\rm cs}(A)_r + (\bar{\lambda}_t, \gamma_{(3)} \lambda_t)\big)$ in which the self-dual part of $\rmd B_s$ is involved, as well as $2B_s \wedge (F_t-\tfrac12\daleth_t, \daleth_t)$, where $B_s$ appears unmodified. 
    
    Given the triviality of the additional fields, one may wonder if their introduction is at all necessary. We note that self-duality of the curvature of $B_s$ follows from the equation of motion of $C_q^+$. Moreover, upon replacing the fields $\beth_s$ and $\daleth_t$ by their on-shell equations, a number of components of the higher gauge potential do not appear in the action: the component $A_p$ of the 1-form potential $A=A_t+A_p$, the antiself-dual part $C_q^-$ of the 3-form potential $C_q$, as well as the 4-form potential $D_v$. One could now simply algebraically fix $C_q^-$ and $D_v$ to ensure the duality equations of motion for the gauge sector, avoiding the requirement for an extension of the action altogether. In this light, it is perhaps not surprising that our construction is possible; it remains, however, non-trivial, which is underlined by the modified supersymmetry transformations~\eqref{eq:shifted_SUSY} as well as the terms in the action that are quadratic in the additional field $\daleth_t$. Moreover, there are a number of reasons why implementing self-duality by algebraic constraints is less than desirable for us. First of all, it is evident that the field content needs to be truncated in some manner, which may contradict an algebraic implementation of the equations of motion. In particular, we note that, from a mathematical perspective, the string Lie 2-algebra (and its metric extension) are only $L_\infty$-subalgebras of the higher gauge algebra we use. The restriction to this natural higher gauge structure requires to impose $C_q=0$, cf.~\cite{Sati:2009ic}, making an algebraic implementation essentially impossible. The precise nature of the truncation will be addressed in future work, but the need for truncating the field content is also seen from the indefinite kinetic term of the tensor multiplet fields. Second, from the perspective of quantization, it certainly does make a difference if the 3-form potential $C_q$ is allowed to have non-self-dual off-shell degrees or not. Moreover, a quantization with a full, unconstrained higher gauge potential can simply follow the canonical BRST prescription and is therefore much more convenient. Third, we are acutely aware that our theory should be generalized in various ways. Recall that the authors of~\cite{Bandos:2013jva} found it necessary to extend the model~\cite{Samtleben:2011fj}, on which our model is based, by the PST mechanism. Similarly, more general higher gauge algebras and more intricate matter couplings should require an actual Sen-like implementation of self-duality. Our model with dynamically implemented self-duality provides a good supersymmetric starting point for this.    
    
    Altogether, we obtain an action that shares many features with an ideal M5-brane model (and significantly differs from it in other regards). Using adjusted metric string structures, its mathematical consistency is guaranteed: the various potential forms do indeed form a connection on a higher principal bundle, which can, in principle, be extended to topologically non-trivial space-times. The action is now accessible to standard perturbative quantization techniques, and we plan to study some quantum aspects in the near future. Clearly, one should quantize  around a background $\langle \phi_s\rangle\gg0$, as otherwise the action is unstable. This is perhaps not surprising, as in a reduction to four dimensions, the expectation value $\langle \phi_s\rangle$ should be related to the Yang--Mills coupling constant (and $\langle B_s\rangle$ should be related to the $\theta$-parameter)~\cite{Saemann:2017zpd,Saemann:2017rjm}. A remaining issue to understand and address in the context of quantization is certainly the indefiniteness of the kinetic term for the tensor multiplet fields. We expect that a truncation to a subset of fields can cure this issue.
    
    We conclude that the two most evident obstacles in constructing an M5-brane model, namely the construction of an appropriate  higher parallel transport and a simple action implementing self-duality, are now fully overcome. However, daunting problems remain, from the choice of the appropriate higher gauge group to formulating interesting conformal interactions of higher gauge fields in a $(2,0)$-supersymmetric form, and it certainly remains unclear if a classical M5-brane model can exist.
    
    \section{Self-dual action from a homotopy algebraic perspective}\label{sec:self-duality_and_homotopy_algebras}
    
    In this section, we study how self-duality is naturally encoded in an action principle from the homotopy algebraic perspective advocated e.g.~in~\cite{Jurco:2018sby}. After a field redefinition, the action equals the one proposed by Sen in previous work~\cite{Sen:2015nph,Sen:2019qit}, see also~\cite{Sen:2015uaa}. In both forms, the action is a useful alternative to the PST mechanism. For definiteness, we shall focus on the case interesting to us, i.e.~a 2-form potential $B$ with self-dual curvature on six-dimensional Minkowski space. This constraint, however, is merely to simplify our discussion, which readily extends to the general case.

    \subsection{Dynamical data as homotopy algebra}
    
    Consider six-dimensional Minkowski space $M=\IR^{1,5}$ with Hodge star operator $\star$ and codifferential $\rmd^\dagger=\star \rmd \star$. Let $B$ be a 2-form potential with curvature
    \begin{subequations}\label{eq:2-form-dynamics}
        \begin{equation}
            H\coloneqq\rmd B
        \end{equation}
        and the obvious higher gauge structure 
        \begin{equation}\label{eq:2-form-gauge}
            B\mapsto \tilde B=B+\rmd \Lambda\eand \Lambda \mapsto \tilde \Lambda=\Lambda+\rmd \lambda
        \end{equation}
        for $\Lambda\in \Omega^1(M)$ and $\lambda\in \Omega^0(M)$. We are interested in the self-duality equation
        \begin{equation}\label{eq:2-form-eom}
            H=\star H
        \end{equation}
        as an equation of motion.
    \end{subequations}
    
    Gauge symmetries are naturally encoded in a BRST differential, which is nothing but a Chevalley--Eilenberg differential, defining the dual of an $L_\infty$-algebra. This $L_\infty$-algebra captures the kinematical data of a gauge theory. As familiar from the BV formalism, one can extend the BRST complex by antifields in order to capture the full dynamics of a gauge theory. The action of the extended differential on the antifields is given by the generators of the on-shell ideal, i.e.~the expressions that vanish by the equations of motion. Dually, again, this simply yields an extension of the $L_\infty$-algebra capturing the kinematical data to one that contains the full dynamical information. In the following, we describe this $L_\infty$-algebra for the dynamical data~\eqref{eq:2-form-dynamics}, reviewing some basic notions of $L_\infty$-algebras.
    
    An {\em $L_\infty$-algebra} is a graded vector space $\sfL=\oplus_{n\in\IZ} \sfL_n$ with totally antisymmetric multilinear {\em higher products} $\mu_i:\sfL^{\wedge i}\rightarrow \sfL$ of degree~$2-i$, which satisfy the {\em homotopy Jacobi identities} 
    \begin{equation}
        \sum_{i+j=n} \sum_{\sigma\in {\rm Sh}(j;i)} (-1)^{j}\chi(\sigma; \ell_1,\ldots,  \ell_n) \mu_{j+1}(\mu_i( \ell_{\sigma(1)},\ldots,v_{\sigma(i)}), \ell_{\sigma(i+1)},\ldots, \ell_{\sigma(n)})=0
    \end{equation}
    for all $\ell_i\in \sfL$. Here, ${\rm Sh}(j;i)$ is the set of unshuffles, i.e.~permutations of $i$ elements preserving the relative order of the first $j$ and the last $i-j$ elements and $\chi$ is the antisymmetric Koszul sign $(-1)^p$, where $p$ is the number of permutations in $\sigma$ involving an even element. 
    
    In an $L_\infty$-algebra, we can regard elements $a\in \sfL_1$ of degree~1 as generalized {\em gauge potentials} with {\em curvature}
    \begin{equation}
        f\coloneqq\mu_1(a)+\tfrac12\mu_2(a,a)+\tfrac1{3!}\mu_3(a,a,a)+\ldots~~\in \sfL_2~.
    \end{equation}
    The {\em homotopy Maurer--Cartan equation} is then simply
    \begin{equation}\label{eq:hMC}
        f=\sum_{i=1}^{\infty} \frac{1}{i!}\mu_i(a,\ldots,a)=0~.
    \end{equation}
    Generalized gauge potentials transform under {\em gauge transformations}, parameterized by elements $c_0\in \sfL_0$ as
    \begin{subequations}
        \begin{equation}\label{eq:hMCGaugeTrafo}
            \delta_{ c_0} a\ \coloneqq\ \sum_{i\geq0} \frac{1}{i!}\mu_{i+1}(a,\ldots,a, c_0)~;
        \end{equation}
        higher gauge transformations of $c_0$ are parameterized by an element $c_{-1}\in \sfL_{-1}$ and given by 
        \begin{equation}\label{eq:hMCHigherGaugeTrafo}
            \delta_{ c_{-1}} c_0\ \coloneqq\ \sum_{i\geq0} \frac{1}{i!}\mu_{i+1}(a,\ldots,a, c_{-1})
        \end{equation}
    \end{subequations}
    with an evident generalization to $c_{-k}\in\sfL_{-k}$, $k\in \IN$. For more details, see e.g.~\cite{Jurco:2018sby}.
    
    The dynamics~\eqref{eq:2-form-dynamics} of a 2-form potential $B$ with self-dual curvature $H$ is described by the homotopy Maurer--Cartan equation of the $L_\infty$-algebra $\sfL^{\rm sd}$ given by the differential chain complex
    \begin{equation}\label{eq:2-form-complex}
        \sfL^{\rm sd}=\left(
        \begin{tikzcd}
            \underbrace{\stackrel{\lambda}{\Omega^0(M)}}_{\sfL^{\rm sd}_{-1}} \arrow[r,"\rmd"] & \underbrace{\stackrel{\Lambda}{\Omega^1(M)}}_{\sfL^{\rm sd}_{0}} \arrow[r,"\rmd"] & \underbrace{\stackrel{B}{\Omega^2(M)}}_{\sfL^{\rm sd}_{1}} \arrow[r,"-(\rmd-\star \rmd)"] &[25pt] \underbrace{\stackrel{H^-}{\Omega^3_-(M)}}_{\sfL^{\rm sd}_{2}}
        \end{tikzcd}
        \right)~.
    \end{equation}
    Here, $\Omega^3_-(M)$ denotes the vector space of antiself-dual 3-forms. The equation of motion~\eqref{eq:2-form-eom} is clearly the homotopy Maurer--Cartan equation~\eqref{eq:hMC} specialized to $\sfL^{\rm sd}$. Similarly, gauge and higher gauge transformations~\eqref{eq:2-form-gauge} are given by specializations of~\eqref{eq:hMCGaugeTrafo} and~\eqref{eq:hMCHigherGaugeTrafo} to $\sfL^{\rm sd}$.
    
    \subsection{Action from homotopy algebras}
    
    In order to have an action principle, we need the analogue of an inner product on the $L_\infty$-algebra encoding our dynamics. An inner product on an $L_\infty$-algebra $\sfL$, which is usually called a {\em cyclic structure} on $\sfL$, is a graded symmetric map $\langle-,-\rangle:\sfL\times \sfL\rightarrow \IR$ satisfying the cyclic permutation condition
    \begin{equation}\label{eq:cyclicity}
        \langle\ell_1,\mu_i(\ell_2,\ldots,\ell_{i+1})\rangle\ =\ (-1)^{i+i(|\ell_1|+|\ell_{i+1}|)+|\ell_{i+1}|\sum_{j=1}^{i}|\ell_j|}\langle\ell_{i+1},\mu_i(\ell_1,\ldots,\ell_{i})\rangle~.
    \end{equation}
    In general, cyclic structures arise from certain symplectic forms on the graded vector space $\sfL$. Together with the differential, this is the data for writing down the classical master equation in the Batalin--Vilkovisky formalism and  the symplectic form is the one giving rise to the Poisson bracket known as the antibracket.
    
    For cyclic $L_\infty$-algebras with cyclic structure $\langle-,-\rangle$ of degree~$-3$, the homotopy Maurer--Cartan equation~\eqref{eq:hMC} is the equation of motion of the action
    \begin{equation}
        S_{\rm hMC}=\sum_{i=1}^\infty\frac{1}{i+1}\langle a,\mu_i(a,\ldots, a)\rangle~.
    \end{equation}
    
    In the case of the $L_\infty$-algebra $\sfL^{\rm sd}$, the dimensionality of the homogeneous parts of the underlying graded vector space shows that there is no non-degenerate pairing of degree~$-3$. However, one can trivially extend an $L_\infty$-algebra $\sfL$ to a cyclic one by doubling it to $T^*[-3]\sfL$. Here $[-3]$ indicates a shift of the cotangent fibers by $3$.\footnote{This is, in fact, precisely what one does when introducing antifields in the BV formalism. The shift here is only by $1$ since the ghost degree counting is the inverse of the $L_\infty$-degree, shifted by $1$. For a detailed explanation, see again~\cite{Jurco:2018sby}.} 
    
    We thus introduce antifields\footnote{We shall employ a $\dagger$ to denote antifields, while $\pm$ will denote self-dual and antiself-dual parts of 3-forms.} $\lambda^\dagger$, $\Lambda^\dagger$, $B^\dagger$, $H^\dagger$ for the ghost-for-ghost, the ghost, the field and the antiself-dual part of the curvature, respectively, and consider the differential complex
    \begin{equation}
        \begin{aligned}
            \hat \sfL^{\rm sd}&=\left(
            \begin{tikzcd}[row sep=0cm,column sep=16.5pt]
                \stackrel{\lambda}{\Omega^0(M)} \arrow[r,"\rmd"] & \stackrel{\Lambda}{\Omega^1(M)} \arrow[r,"\rmd"] & \stackrel{B}{\Omega^2(M)} \arrow[r,"-(\rmd-\star \rmd)"]  &[24pt] \stackrel{H^-}{\Omega^3_- (M)} & 
                \\
                & & \oplus & \oplus
                \\
                \underbrace{\phantom{\stackrel{\lambda}{\Omega^0(M)}}}_{\hat \sfL^{\rm sd}_{-1}} & \underbrace{\phantom{\stackrel{\lambda}{\Omega^0(M)}}}_{\hat \sfL^{\rm sd}_{0}} & \underbrace{\stackrel{H_-^\dagger}{\Omega^3_+(M)}}_{\hat \sfL^{\rm sd}_{1}} \arrow[r,"\rmd^\dagger"] & \underbrace{\stackrel{B^\dagger}{\Omega^2(M)}}_{\hat \sfL^{\rm sd}_{2}}\arrow[r,"\rmd^\dagger"]  & \underbrace{\stackrel{\Lambda^\dagger}{\Omega^1(M)}}_{\hat \sfL^{\rm sd}_{3}} \arrow[r,"\rmd^\dagger"] & \underbrace{\stackrel{\lambda^\dagger}{\Omega^0(M)}}_{\hat \sfL^{\rm sd}_{4}}
            \end{tikzcd}
            \right).
        \end{aligned}
    \end{equation}
    Note that in six dimensions, self-dual forms naturally pair with antiself-dual forms and vice versa; $H_-^\dagger$ is self-dual for this reason.
    
    The diagonal inner product pairs the fields in an obvious way with a sign $(-1)^{|\Phi|}$ for each field $\Phi$:
    \begin{equation}
        \begin{aligned}
            \langle~&\lambda_1+\lambda_1^\dagger+\Lambda_1+\Lambda_1^\dagger+\ldots~,~\lambda_2+\lambda_2^\dagger+\Lambda_2+\Lambda_2^\dagger+\ldots~\rangle\\
            &\hspace{1cm}=\int_M-\lambda_1\wedge \star\lambda^\dagger_2-\lambda_2\wedge \star\lambda^\dagger_1+\Lambda_1\wedge \star\Lambda_2^\dagger+\Lambda_2\wedge \star\Lambda_1^\dagger-\ldots~.
        \end{aligned}
    \end{equation}
    A gauge potential of this $L_\infty$-algebra, i.e.~an element of $\hat \sfL^{\rm sd}_1$, is then a 2-form $B$ together with a self-dual 3-form $H_-^\dagger$. The homotopy Maurer--Cartan action of the cyclic $L_\infty$-algebra $\hat \sfL^{\rm sd}$ reads as
    \begin{equation}
        S_{\rm hMC}=\int_M H_-^\dagger\wedge \star (\rmd-\star \rmd)B~,
    \end{equation}
    and variation with respect to $H_-^\dagger$ leads to the expected equation of motion $H=\star H$. Varying with respect to $B$ yields the equation of motion $\rmd^\dagger H_-^\dagger=0$, which is equivalent to $\rmd H_-^\dagger=0$.
    
    Let us now rotate the field content by performing the following field redefinition:
    \begin{equation}\label{eq:relabel}
        \begin{aligned}
            \beth&\coloneqq H^\dagger_-+(\rmd+\star \rmd) B~,~~~&\beth^\dagger&\coloneqq H^-~,\\
            B&\rightarrow B~,~~~&B^\dagger&\rightarrow B^\dagger+2\star \rmd^\dagger \beth^\dagger~,
        \end{aligned}
    \end{equation}
    where $\beth\in \Omega^3_+$ is a self-dual 3-form and $\beth^\dagger\in \Omega^3_-$ is an antiself-dual 3-form. Because of
    \begin{equation}
        \int_M((\rmd+\star \rmd) B)\wedge \star \beth^\dagger=\int_M2B\wedge \star \rmd^\dagger \beth^\dagger~,
    \end{equation}
    the inner product remains diagonal:
    \begin{equation}
        \begin{aligned}
            \langle~\ldots+B_1+\beth_1+&\beth_1^\dagger+B_1^\dagger\dots~,~\dots+B_2+\beth_2+\beth_2^\dagger+B_2^\dagger+\ldots~~\rangle\\
            &=\int_M-B_1\wedge \star B_2^\dagger-B_2\wedge \star B_1^\dagger-\beth_1\wedge \star \beth_2^\dagger-\beth_2\wedge \star \beth_1^\dagger+\ldots~.
        \end{aligned}
    \end{equation}
    After the field redefinition, the differential graded complex reads as
    \begin{equation}
        \begin{aligned}
            \tilde \sfL^{\rm sd}=\left(
            \begin{tikzcd}[row sep=0cm,column sep=21pt]
                \stackrel{\lambda}{\Omega^0(M)} \arrow[r,"\rmd"] & \stackrel{\Lambda}{\Omega^1(M)} \arrow[r,"\rmd"] & \stackrel{B}{\Omega^2(M)} \arrow[r,"\mu_1"] \arrow[start anchor=south east, end anchor= north west,ddr] & \stackrel{B^\dagger}{\Omega^2 (M)} \arrow[r,"\rmd^\dagger"] & \stackrel{\Lambda^\dagger}{\Omega^1(M)} \arrow[r,"\rmd^\dagger"] & \stackrel{\lambda^\dagger}{\Omega^0(M)}
                \\
                & & \oplus & \oplus 
                \\
                \underbrace{\phantom{\stackrel{\lambda}{\Omega^0(M)}}}_{\tilde \sfL^{\rm sd}_{-1}} & \underbrace{\phantom{\stackrel{\lambda}{\Omega^0(M)}}}_{\tilde \sfL^{\rm sd}_{0}} & \underbrace{\stackrel{\beth}{\Omega^3_+(M)}}_{\tilde \sfL^{\rm sd}_{1}} \arrow[start anchor=north east, end anchor= south west,crossing over,uur,swap] & \underbrace{\stackrel{\beth^\dagger}{\Omega^3_-(M)}}_{\tilde \sfL^{\rm sd}_{2}} & \underbrace{\phantom{\stackrel{\lambda}{\Omega^0(M)}}}_{\tilde \sfL^{\rm sd}_{3}} & \underbrace{\phantom{\stackrel{\lambda}{\Omega^0(M)}}}_{\tilde \sfL^{\rm sd}_{4}}
            \end{tikzcd}
            \right)~,
        \end{aligned}
    \end{equation}
    where the differential $\mu_1$ is given by 
    \begin{equation}
        \mu_1(B,\beth)=\big(-\rmd^\dagger \rmd B-\rmd^\dagger \beth~,~-(\rmd-\star \rmd )B\big)~.
    \end{equation}
    The homotopy Maurer--Cartan action of this $L_\infty$-algebra is then 
    \begin{equation}
        S=\int \Big(\tfrac12 \rmd B\wedge \star\rmd B-\rmd B\wedge \beth\Big)~,
    \end{equation}
    where we once integrated by parts. Coupling this to potential matter fields $\Psi$ exclusively through $\beth$, we obtain Sen's action,
    \begin{equation}\label{eq:sen_action}
        S_{\rm Sen}=\int \Big(\tfrac12 \rmd B\wedge \star\rmd B-\rmd B\wedge \beth+\caL_{\rm int}(\beth,\Psi)\Big)~,
    \end{equation}
    cf.~\cite{Sen:2015nph,Sen:2019qit,Sen:2015uaa}. Its equations of motion read as 
    \begin{equation}\label{eq:sen_eom}
        \rmd(\star \rmd B-\beth)=0~,~~~\rmd B-\star \rmd B+\frac{\delta \int \caL_{\rm int}(\beth,\Psi)}{\delta \beth}=0~,~~~\frac{\delta \int \caL_{\rm int}(\beth,\Psi)}{\delta \Psi}=0~.
    \end{equation}
    The second equation of motion determines the antiself-dual part of the curvature $H=\rmd B$ by a variation of $\caL_{\rm int}(\beth,\Psi)$. The self-dual part $H^+=\tfrac12(H+\star H)$ linearly combines with $\beth$ into the two expressions
    \begin{equation}\label{eq:3-forms_Sen}
        2H^+-\beth\eand 2H^++\beth~.
    \end{equation}
    We shall give the physical interpretation of these two 3-forms in the next section.
    
    \subsection{Physical interpretation of the fields}
    
    In Sen's papers, the 3-form $2H^++\beth$ is regarded as a physical degree of freedom, while the 3-form $2H^+-\beth$ is considered unphysical. It is free by the first equation of motion in~\eqref{eq:sen_eom} and it comes with the wrong sign kinetic term. A detailed analysis then shows that it decouples from the theory~\cite{Sen:2019qit}. In the homotopy algebraic picture, we can simply identify the field $2H^+-\beth$ as pure gauge, as we shall show in the following.
    
    Above, we used the fact that the relabeling of fields~\eqref{eq:relabel} is an automorphism on $\hat \sfL^{\rm sd}$ with an underlying chain isomorphism and such maps link $L_\infty$-algebras whose corresponding field theories are evidently equivalent. Besides field redefinitions, equivalent field theories can be linked by integrating in and out additional fields. It is therefore clear that equivalence of field theories\footnote{at the classical level; equivalence of quantum field theories is more involved, cf.~e.g.~\cite{Borsten:2020zgj}} corresponds to a wider notion of isomorphism. As explained in~\cite{Jurco:2018sby}, the appropriate notion is that of a {\em quasi-isomorphism} of $L_\infty$-algebra, which is a general morphism of $L_\infty$-algebra which descends to isomorphisms between the cohomologies of the differential complexes contained in the $L_\infty$-algebras. Quasi-isomorphic $L_\infty$-algebras with cyclic structure of degree~$-3$ then correspond to classical field theories with the same\footnote{up to field redefinitions} tree-level scattering amplitudes~\cite{Macrelli:2019afx}. 
    
    In the case at hand of the $L_\infty$-algebras $\sfL^{\rm sd}$, $\hat \sfL^{\rm sd}$, and $\tilde \sfL^{\rm sd}$, which are differential complexes, a quasi-isomorphism of $L_\infty$-algebras is the same as a quasi-isomorphism of differential complexes, i.e.~a chain map which descends to an isomorphism on the cohomologies. Clearly, the $L_\infty$-algebras $\sfL^{\rm sd}$ and $\hat \sfL^{\rm sd}\cong \tilde \sfL^{\rm sd}$ have different cohomologies, and they are not fully physically equivalent. 
    
    The $L_\infty$-algebra $\tilde \sfL^{\rm sd}$, however, allows for an extension by introducing a gauge symmetry for $H_-^\dagger$, 
    \begin{equation}
        H_-^\dagger\rightarrow \tilde H_-^\dagger=H_-^\dagger+\rmd \alpha~,~~~\alpha\in \Omega^2_+\coloneqq\Omega^2\cap\ker(\rmd-\star\rmd)~,
    \end{equation}
    which fits into the following cyclically extended differential complex:
    \begin{equation}
        \left(
        \begin{tikzcd}[row sep=0cm,column sep=21pt]
            \stackrel{\lambda}{\Omega^0(M)} \arrow[r,"\rmd"] & \stackrel{\Lambda}{\Omega^1(M)} \arrow[r,"\rmd"] & \stackrel{B}{\Omega^2(M)} \arrow[r,"-(\rmd-\star \rmd)"]  &[25pt] \stackrel{H^-}{\Omega^3_- (M)} \arrow[r,"\delta"]  & \stackrel{\alpha^\dagger}{\Omega^2_+(M)}
                \\
                & \oplus & \oplus & \oplus & \oplus
                \\
            & \stackrel{\alpha}{\Omega^2_+(M)} \arrow[r,"\rmd"]  & \stackrel{H_-^\dagger}{\Omega^3_+(M)} \arrow[r,"\rmd^\dagger"] & \stackrel{B^\dagger}{\Omega^2(M)}\arrow[r,"\rmd^\dagger"]  & \stackrel{\Lambda^\dagger}{\Omega^1(M)} \arrow[r,"\rmd^\dagger"] & \stackrel{\lambda^\dagger}{\Omega^0(M)}
        \end{tikzcd}
        \right)~.
    \end{equation}
    Here, the operator $\delta$ is given by the adjoint of $\rmd:\Omega^2_+(M)\rightarrow \Omega^3_+(M)$:
    \begin{equation}
        \langle \beth^\dagger,\rmd \alpha\rangle=\langle \delta \beth^\dagger,\alpha\rangle~.
    \end{equation}

    We note that the complex is exact at $\Omega^3_+(M)$ and $\Omega^3_-(M)$:
    \begin{equation}
        \rmd^\dagger H_-^\dagger=0~~\Leftrightarrow~~\rmd H_-^\dagger=0~~\Leftrightarrow ~~H_-^\dagger=\rmd \alpha~,
    \end{equation}
    where the last equivalence is Poincar\'e's lemma. Thus, the field $H_-^\dagger$ does not give rise to an observable. This is an even  stronger statement than the statement that $H_-^\dagger=\beth-2H^+$ decouples.
    
    After the field redefinition~\eqref{eq:relabel}, the gauge transformation takes the form
    \begin{equation}
        \beth\rightarrow \tilde \beth=\beth+\rmd \alpha~,~~~\alpha\in \Omega^2_+\coloneqq\Omega^2\cap\ker(\rmd-\star\rmd)~,
    \end{equation}
    and the cyclically extended complex $\tilde \sfL^{\rm sdg}$ is 
    \begin{equation}
        \tilde \sfL^{\rm sdg}_\alpha=\left(
        \begin{tikzcd}[row sep=0cm,column sep=21pt]
            \stackrel{\lambda}{\Omega^0(M)} \arrow[r,"\rmd"] & \stackrel{\Lambda}{\Omega^1(M)} \arrow[r,"\rmd"] & \stackrel{B}{\Omega^2(M)} \arrow[r,"\mu_1"] \arrow[start anchor=south east, end anchor= north west,ddr] & \stackrel{B^\dagger}{\Omega^2 (M)} \arrow[r,"\rmd^\dagger"] & \stackrel{\Lambda^+}{\Omega^1(M)} \arrow[r,"\rmd^\dagger"] & \stackrel{\lambda^\dagger}{\Omega^0(M)}
                \\
                & \oplus& \oplus & \oplus & \oplus
                \\
            \underbrace{\phantom{\stackrel{\lambda}{\Omega^0(M)}}}_{\tilde \sfL^{\rm sd}_{-1}}& \underbrace{\stackrel{\alpha}{\Omega^2_+(M)}}_{\tilde \sfL^{\rm sd}_{0}} \arrow[r,"\rmd"] & \underbrace{\stackrel{\beth}{\Omega^3_+(M)}}_{\tilde \sfL^{\rm sd}_{1}} \arrow[start anchor=north east, end anchor= south west,crossing over,uur,swap] & \underbrace{\stackrel{\beth^\dagger}{\Omega^3_-(M)}}_{\tilde \sfL^{\rm sd}_{2}} \arrow[r,"\delta"]  & \underbrace{\stackrel{\alpha^\dagger}{\Omega^2_+(M)}}_{\tilde \sfL^{\rm sd}_{3}} & \underbrace{\phantom{\stackrel{\lambda}{\Omega^0(M)}}}_{\tilde \sfL^{\rm sd}_{4}}
        \end{tikzcd}
        \right).
    \end{equation}

    We note that our construction is not particularly sophisticated. It is merely the evident observation that the cohomologies at the end of chain complexes can be rendered trivial by adding an injection into the kernel or a surjection out of the cokernel of the differential. The interesting point is that this can be generalized from chain complexes to $L_\infty$-algebras, see~\cite[Proposition C.1]{Saemann:2019dsl}.
    
    \subsection{Physical interpretation in the presence of matter fields}
    
    Let us also briefly comment on the situation of non-trivial couplings to matter fields as in the action~\eqref{eq:sen_action}. If the terms in $\caL_{\rm int}(\beth,\Psi)$ involving simultaneously $\beth$ and matter fields are at least of cubic order, they only enter into higher products $\mu_i$ with $i\geq 2$ and do not distort the differential complex, which thus reads
    \begin{equation}
        \left(
        \begin{tikzcd}[row sep=0cm,column sep=18pt]
            & & \stackrel{\Psi}{\Gamma(E)} \arrow[r,"{\rm kin}"] &[25pt] \stackrel{\Psi^\dagger}{\Gamma(E^*)} 
                \\
                & & \oplus & \oplus & 
                \\
            \stackrel{\lambda}{\Omega^0(M)} \arrow[r,"\rmd"] & \stackrel{\Lambda}{\Omega^1(M)} \arrow[r,"\rmd"] & \stackrel{B}{\Omega^2(M)} \arrow[r,"-(\rmd-\star \rmd)"] & \stackrel{H^-}{\Omega^3_- (M)} \arrow[r,"\delta"]  & \Omega^2_+(M) 
                \\
                & \oplus & \oplus & \oplus & \oplus
                \\
            & \Omega^2_+(M) \arrow[r,"\rmd"] & \stackrel{H_-^\dagger}{\Omega^3_+(M)} \arrow[r,"\rmd^\dagger"] & \stackrel{B^\dagger}{\Omega^2(M)}\arrow[r,"\rmd^\dagger"]  & \stackrel{\Lambda^\dagger}{\Omega^1(M)} \arrow[r,"\rmd^\dagger"] & \stackrel{\lambda^\dagger}{\Omega^0(M)}
        \end{tikzcd}
        \right),
    \end{equation}
    where $E$ is some vector bundle whose sections $\Gamma(E)$ form the field space for the matter fields $\Psi$ and `${\rm kin}$' denotes their kinetic operator. 
    
    In order to lift elements in $\Omega^2_+$ to gauge parameters and implement the gauge symmetry of the full interacting theory, one needs to enhance the kernel injection $\Omega^2_+(M)\xhookrightarrow{~\rmd~}\Omega^3_+(M)$ from the level of chain complexes to the level of $L_\infty$-algebras, as explained in~\cite[Proposition C.1]{Saemann:2019dsl}. This covariantizes the abelian gauge symmetry $H_-^\dagger\rightarrow \tilde H_-^\dagger=H_-^\dagger+\rmd \alpha$ with $\alpha\in \Omega^2_+$. After this, the complex is still exact at $\Omega^3_+(M)$ and our above analysis is still valid.
    
    We can also discuss the case\footnote{This case was not treated in~\cite{Sen:2019qit}.} in which $\caL_{\rm int}(\beth,\Psi)$ does contain quadratic terms of the form $\langle\beth,D\Psi\rangle$, where $D$ is some operator, possibly a differential operator. Here, the differential complex mixes with the one for the matter fields, and we get the complex
    \begin{equation}
        \left(
        \begin{tikzcd}[row sep=0cm,column sep=18pt]
            & \Omega^2_r(M) \arrow[r,"\rmd"] & \stackrel{H_-^\dagger}{\Omega^3_+(M)} \arrow[r,"\rmd^\dagger"] \arrow[start anchor=south east, end anchor= north west,ddr,"D" {pos=0.15}]&[25pt] \stackrel{B^\dagger}{\Omega^2(M)}\arrow[r,"\rmd^\dagger"]  & \stackrel{\Lambda^\dagger}{\Omega^1(M)} \arrow[r,"\rmd^\dagger"] & \stackrel{\lambda^\dagger}{\Omega^0(M)}
                \\
                & & \oplus & \oplus & 
                \\
            & \oplus & \stackrel{\Psi}{\Gamma(E)} \arrow[start anchor=north east, end anchor= south west,crossing over,uur,swap,dashed] \arrow[start anchor=south east, end anchor= north west,ddr,dashed] \arrow[r,"{\rm kin}"] & \stackrel{\Psi^\dagger}{\Gamma(E^*)} & \oplus
                \\
                & & \oplus & \oplus & 
                \\
            \stackrel{\lambda}{\Omega^0(M)} \arrow[r,"\rmd"] & \stackrel{\Lambda}{\Omega^1(M)} \arrow[r,"\rmd"] & \stackrel{B}{\Omega^2(M)}\arrow[start anchor=north east, end anchor= south west,crossing over,uur,dashed]  \arrow[r,"-(\rmd-\star \rmd)"]  & \stackrel{H^-}{\Omega^3_- (M)} \arrow[r,"\delta"]  & \Omega^2_r(M)
        \end{tikzcd}
        \right)~,
    \end{equation}
    where the operators on the dashed arrows are irrelevant for our discussion. In order to recover a complex, we can simply define
    \begin{equation}
        \Omega^2_r\coloneqq\Omega^2_+\cap \ker(D\rmd)~.
    \end{equation}
    Again, for an interacting theory, one should covariantize the gauge symmetries using~\cite[Proposition C.1]{Saemann:2019dsl}.  Otherwise, our above analysis is still valid: on-shell, $H_-^\dagger$ is still exact by the Poincar\'e lemma and thus there are no new observables. 
    
    Altogether, we have shown that the observables of the action~\eqref{eq:sen_action} arise from a 2-form potential $B$ with self-dual curvature and additional matter fields. The field combination $H_-^\dagger\coloneqq\beth-(\rmd+\star \rmd)B=\beth-2H^+$ does not enter the observables.
    
    \section{The six-dimensional $\caN=(1,0)$ supersymmetric field theory}\label{sec:old_field_theory}
    
    In this section, we briefly review the model presented in~\cite{Saemann:2017rjm,Saemann:2017zpd}, which we will improve in the next section. We start with a review of the gauge structure, before we present the Lagrangian, its equations of motion and supersymmetry transformations.
    
    \subsection{Higher gauge algebra: metric string structures}
    
    A crucial feature of six-dimensional superconformal field theories is the presence of a 2-form gauge potential. Just as ordinary gauge potentials are mathematically connections on principal fiber bundles, such 2-form gauge potentials are really part of connections on categorified principal fiber bundles also known as {\em gerbes}. Abelian gerbes are well-understood and appear in various contexts. Non-abelian gerbes have been defined~\cite{Breen:math0106083,Aschieri:2003mw}, but their connections, as usually defined, are locally gauge equivalent to abelian ones~\cite{Gastel:2018joi,Saemann:2019dsl}. This makes the construction of local, non-trivially interacting Lagrangians essentially impossible.
    
    There is a particular class of higher principal bundles called {\em string structures}, which do not suffer from this problem~\cite{Killingback:1986rd,Redden:2006aa,Sati:2008eg,Waldorf:2009uf}. They encode spin structures on the loop space of their base manifold and arise naturally within string theory and supergravity. Their higher structure group allows for a deformed or adjusted notion of curvature which induces also a modified non-abelian gauge transformation on the 2-form gauge potential. To construct a Lagrangian, however, we need to extend these to metric string structures as first done in~\cite{Saemann:2017rjm} and discussed in detail in~\cite{Saemann:2019dsl}. In the following, we just review the bare minimum of the underlying algebraic structure necessary for the construction of our field theory.
    
    Given a Lie algebra $\frg$ endowed with a metric $(-,-)$, the metric extension\footnote{This is, in fact, the skeletal model of a cyclic Lie 3-algebra, minimally extended to a Lie 4-algebra.}  $\aghsk^\omega$ is given by an $L_\infty$-algebra with the following underlying complex:
    \begin{equation}
        \aghsk^\omega=\left(
        \begin{tikzcd}[row sep=0cm,column sep=2cm]
            \frg^*_v\arrow[r]{}{\mu_1=\sfid} & \frg^*_u & \IR^*_s \arrow[r]{}{\mu_1=\sfid} & \IR_p^*
            \\
            & \oplus & \oplus & \oplus
            \\
            \underbrace{\phantom{\frg^*_v}}_{\hat \frg^\omega_{\rm sk,-3}} & \underbrace{\IR_q}_{\hat \frg^\omega_{\rm sk,-2}} \arrow[r]{}{\mu_1=\sfid} & \underbrace{\IR_r}_{\hat \frg^\omega_{\rm sk,-1}} & \underbrace{\frg_t}_{\hat \frg^\omega_{\rm sk,0}}
        \end{tikzcd}\right)~,
    \end{equation}
    where the subscripts merely help to identify different copies of the same vector space: $\frg_v^*\cong \frg_u^*\cong \frg^*$, $\frg_t\cong \frg$ etc. The map $\mu_1$ acts as indicated and trivially on all other vector spaces. It thus satisfies $\mu_1^2=0$. We also have the following higher products:
    \begin{equation}\label{eq:brackets_skeletal_metric}
        \begin{aligned}
            &\mu_2:\frg_t\wedge \frg_t\rightarrow \frg_t~,~~~&&\mu_2(t_1,t_2)=[t_1,t_2]~,\\
            &\mu_2:\frg_t\wedge\frg^*_u\rightarrow \frg^*_u~,~~~&&\mu_2(t,u)=u\big([-,t]\big)~,\\
            &\mu_2:\frg_t\wedge\frg_v^*\rightarrow \frg_v^*~,~~~&&\mu_2(t,v)=v\big([-,t]\big)~,\\
            &\mu_3:\frg_t\wedge \frg_t \wedge \frg_t\rightarrow \IR_r~,~~~&&\mu_3(t_1,t_2,t_3)=(t_1,[t_2,t_3])~,\\
            &\mu_3:\frg_t\wedge \frg_t \otimes\IR^*_s\rightarrow \frg^*_u~,~~~&&\mu_3(t_1,t_2,s)= s\big(\,(-,[t_1,t_2])\,\big)~.
        \end{aligned}
    \end{equation}
    As explained in~\cite{Borsten:2021ljb}, the underlying algebraic structure is in fact a generalized notion of $L_\infty$-algebra known as an $EL_\infty$-algebra and as a remnant of this structure, we have the additional maps
    \begin{equation}
        \begin{aligned}
            &\nu_2:\frg_t \odot \frg_t \rightarrow \IR_r~,~~~&&\nu_2(t_1,t_2)=(t_1,t_2)~,
            \\
            &\nu_2: \frg_t \odot \IR^*_s \rightarrow \frg^*_u ~,~~~&&\nu_2(t,s)=2s(-,t)~,
            \\
            &\nu_2:\frg_t \odot \frg^*_u\rightarrow \frg^*_v~,~~~&&\nu_2(t_1,u_1)=u_1\big([-,t_1]\big)~,
            \\
            &\nu_3:\frg_t \wedge \frg_t\otimes\IR^*_s\rightarrow \frg^*_v~,~~~&&\nu_3(t_1,t_2,s)=s\big(-,[t_1,t_2]\big)~.
        \end{aligned}
    \end{equation}
    The precise details of $EL_\infty$-algebras are irrelevant for our purposes; note, however, that the fact that $\aghsk^\omega$ is an $EL_\infty$-algebra is crucial in constructing the appropriate finite gauge transformation as well as higher principal bundles with $\aghsk^\omega$ as a structure $EL_\infty$-algebra. Since $EL_\infty$-algebras are good higher versions of Lie algebras, abstract nonsense guarantees the existence and mathematical consistency of such higher bundles.

    We also define the following two auxiliary maps, which are not part of the $EL_\infty$-structure:
    \begin{equation}\label{eq:aux_nu}
        \begin{aligned}
            &\hat \nu_2:\frg_t \odot \frg_t \rightarrow \IR_q~,~~~&&\hat \nu_2(t_1,t_2)=(t_1,t_2)~,
            \\
            &\hat \nu_2: \frg_t \odot \IR^*_s \rightarrow \frg^*_v ~,~~~&&\hat \nu_2(t,s)=2s(-,t)~.
        \end{aligned}
    \end{equation}
    Finally, there are the evident pairings
    \begin{equation}\label{eq:pairings}
        \begin{aligned}
            &\langle -,-\rangle: \frg^*_u\times \frg_t \rightarrow \IR~,~~~&&\langle u,t\rangle=u(t)~,
            \\
            &\langle -,-\rangle: \IR_s^*\times \IR_r\rightarrow \IR~,~~~&&\langle s,r\rangle=s(r)~,
            \\
            &\langle -,-\rangle: \IR_p^*\times \IR_q\rightarrow \IR~,~~~&&\langle p,q\rangle=p(q)~.
        \end{aligned}
    \end{equation}
    There is no pairing for elements in $\frg_v^*$, but fields taking values in $\frg_v^*$ will only appear as arguments of $\mu_1$ in the action, which maps them to $\frg_u^*$.
    
    In order to extend the pairing $\langle-,-\rangle$ to a cyclic structure, cf.~\eqref{eq:cyclicity}, we need to extend it in a graded symmetric way to all of $\aghsk^\omega$. Such a graded symmetric structure would be very natural in a topological field theory, in which we may want to define non-vanishing graded expressions such as $\langle H,H\rangle$, where $H$ is a 3-form valued in $\IR_r\oplus \IR_s^*$. In the model we shall consider below, however, we need to incorporate non-vanishing terms of the form $\langle H, \star H\rangle$. These clearly require $\langle-,-\rangle$ to be graded antisymmetric on the subspace $\IR_r\oplus \IR_s^*$. We thus deviate at this point from the metric string structures defined in~\cite{Saemann:2019dsl} and consider only the pairing~\eqref{eq:pairings} without any further symmetrization or antisymmetrization.
    
    Cyclicity of an $L_\infty$-algebra can be replaced by the derivation property
    \begin{equation}\label{eq:cyclic}
        \langle \mu_i(\ell_1,\ldots,\ell_{i-1},\ell_i),\ell_{i+1}\rangle=(-1)^{|\ell_i|\,(i+|\ell_1|+\ldots+|\ell_{i-1}|)}\langle \ell_i,\mu_i(\ell_1,\ldots,\ell_{i-1},\ell_{i+1})\rangle~,
    \end{equation}
    which, for graded symmetric pairings, is equivalent to~\eqref{eq:cyclicity}. This property is readily verified to hold for the higher products introduced above. Moreover, together with similar relations for $\nu_2$, it ensures gauge invariance for actions constructed using this pairing.
    
    \subsection{Adjusted kinematical data}
    
    The local description of an adjusted connection on a metric string structure~\cite{Saemann:2017rjm,Saemann:2017zpd,Saemann:2019dsl} consists of gauge potential forms
    \begin{equation}\label{eq:gauge_potentials_undeformed}
        \begin{aligned}
            A&\in\Omega^1(M)\otimes (\frg_t\oplus \IR^*_p)~,~~~&B&\in \Omega^2(M)\otimes (\IR_r\oplus \IR^*_s)~,\\
            \tilde C&\in \Omega^3(M)\otimes (\frg^*_u\oplus\IR_q)~,~~~&\tilde D&\in \Omega^4(M)\otimes \frg^*_v~,
        \end{aligned}
    \end{equation}
    and the corresponding curvatures are defined as 
    \begin{equation}\label{eq:higher_curvatures_undeformed}
        \begin{aligned}
            F&=\rmd A+\tfrac12 \mu_2(A,A)+\mu_1(B)&&\in~\Omega^2(M)\otimes (\frg_t\oplus \IR^*_p)~,\\
            H&=\rmd B-\tfrac{1}{3!}\mu_3(A,A,A)+\nu_2(A,F)-\mu_1(\tilde C)\\
            &=\rmd B+(A,\rmd A)+\tfrac13(A,\mu_2(A,A))-\mu_1(\tilde C)&&\in~\Omega^3(M)\otimes (\IR_r\oplus \IR^*_s)~,\\
            G&=\rmd \tilde C+\mu_2(A,\tilde C)+\tfrac12\mu_3(A,A,B)+\mu_1(\tilde D)&&\in~\Omega^4(M)\otimes (\frg^*_u\oplus\IR_q)~,\\
            \tilde I&=\rmd \tilde D+\mu_2(A,\tilde D)+\nu_2(F,\tilde C)+\tfrac12 \nu_3(A,A,H)&&\\
            &\phantom{{}={}}+\nu_3(F,A,B)&&\in~\Omega^5(M)\otimes \frg^*_v~.
        \end{aligned}
    \end{equation}
    It will be helpful to have the explicit components of the curvature forms at hand:
    \begin{equation}\label{eq:higher_curvatures_untilded}
        \begin{aligned}
            \begin{pmatrix}
                F_t \\
                F_p
            \end{pmatrix} &= 
            \begin{pmatrix*}[l]
                \rmd A_t + \tfrac12 [A_t,A_t] \\ 
                \rmd A_p + B_s
            \end{pmatrix*}~, 
            \\
            \begin{pmatrix}
                H_r \\
                H_s
            \end{pmatrix} &=
            \begin{pmatrix*}[l]
                \rmd B_r + (A_t, \rmd A_t) + \tfrac13 (A_t,[A_t,A_t]) - \tilde C_q \\
                \rmd B_s
            \end{pmatrix*}~,
            \\
            \begin{pmatrix}
                G_q \\
                G_u
            \end{pmatrix} &=
            \begin{pmatrix*}[l]
                \rmd \tilde C_q \\
                \rmd \tilde C_u + \tilde C_u \wedge ([A_t,-]) + \tfrac12 B_s \wedge ([A_t,A_t],-) + \tilde D_v
            \end{pmatrix*}~,
            \\
            \tilde I_v &= \rmd \tilde D_v - \tilde D_v \wedge ([A_t,-]) - \tilde C_u \wedge ([F_t,-]) + \tfrac12 H_s \wedge ([A_t,A_t],-)\\
            &\hspace{1cm}+ B_s \wedge ([F_t, A_t],-)~,
        \end{aligned} 	
    \end{equation}
    where subscripts indicate again the subspaces of $\aghsk^\omega$ in which the various forms take values.
    
    As one may readily check, these curvatures satisfy the Bianchi identities
    \begin{equation}\label{eq:bianchi_undeformed}
        \begin{aligned}
            \rmd F+\mu_2(A,F)-\mu_1(H)&=0~,~~~&\rmd H-\nu_2(F,F)+\mu_1(G)&=0~,
            \\
            \rmd G+\mu_2(A,G)-\mu_1(\tilde I)&=0~,~~~&\rmd \tilde I+\mu_2(A,\tilde I)-\nu_2(F,G)&=0~.
        \end{aligned}
    \end{equation}
    Infinitesimal gauge transformations are parameterized by differential forms 
    \begin{equation}
        \begin{aligned}
            \Lambda_0&\in \Omega^0(M)\otimes(\frg_t\oplus \IR^*_p)~,~~~&\Lambda_1&\in\Omega^1(M)\otimes(\IR_r\oplus \IR^*_s)~,
            \\
            \Lambda_2&\in \Omega^2(M)\otimes(\frg^*_u\oplus\IR_q)~,~~~&\Lambda_3&\in\Omega^3(M)\otimes\frg^*_v~,
        \end{aligned}
    \end{equation}
    and their actions\footnote{This action is fully fixed by the definition of the curvatures, because gauge transformations are partially flat homotopies, cf.~\cite{Saemann:2019dsl} for details.} on the fields read as
    \begin{subequations}\label{eq:gauge_trafos_undeformed}
        \begin{equation}
            \begin{aligned}
                \delta A&=\rmd \Lambda_0+\mu_2(A,\Lambda_0)-\mu_1(\Lambda_1)~,
                \\
                \delta B&=\rmd \Lambda_1-\nu_2(F,\Lambda_0)+\tfrac12 \mu_3(A,A,\Lambda_0)+\mu_1(\Lambda_2)~,
                \\
                \delta \tilde C&=\rmd \Lambda_2+\mu_2(A,\Lambda_2)+\mu_2(\tilde C,\Lambda_0)-\tfrac12 \mu_3(A,A,\Lambda_1)-\mu_3(A,B,\Lambda_0)-\mu_1(\Lambda_3)~,
                \\
                \delta \tilde D&=\rmd \Lambda_3+\mu_2(A,\Lambda_3)+\mu_2(\tilde D,\Lambda_0)-\nu_2(F,\Lambda_2)-\nu_3(A,F,\Lambda_1)
                \\
                &\hspace{1cm}-\nu_3(A,H,\Lambda_0)-\nu_3(B,F,\Lambda_0)~,
            \end{aligned}
        \end{equation}
        which implies the following actions on the curvature forms:
        \begin{equation}
            \begin{aligned}
                \delta F&=\mu_2(F,\Lambda_0)~,~~~&\delta H&=0~,
                \\
                \delta G&=\mu_2(G,\Lambda_0)~,~~~&\delta \tilde I&=\mu_2(\tilde I,\Lambda_0)~.
            \end{aligned}
        \end{equation}
    \end{subequations}
    This completes the description of adjusted metric string connections.
    
    \subsection{Alternative adjustment}
    
    As explained in~\cite{Saemann:2019dsl}, the form of adjustment is not necessarily unique. It will turn out that supersymmetry and the explicit form of the field theory in which we are interested prefer a slightly different adjustment, which is obtained by the field redefinition 
    \begin{equation}
        C=\tilde C+\tfrac12 \nu_2(A,B)\eand D=\tilde D-\tfrac32 \hat \nu_2(F,B)+\tfrac12 \hat \nu_2(A,H)~,
    \end{equation}
    where $\hat \nu_2$ is the map defined in~\eqref{eq:aux_nu}, as well as a redefinition of curvature
    \begin{equation}
        I=\tilde I-\hat \nu_2(F,H)~.
    \end{equation}
    Note that a redefinition of curvature forms involving exclusively lower curvature forms is harmless as the gauge structure is left invariant. The result is precisely the gauge structure employed in the papers~\cite{Samtleben:2011fj,Samtleben:2012fb,Bandos:2013jva}, cf.~appendix~\ref{app:dictionary} for a dictionary of the gauge algebra structures. To be notationally close to the supergravity literature, we will work with these connections in the following.
    
    The new expressions for the curvatures are 
    \begin{equation}
        \begin{aligned}
            G&=\rmd C + \mu_2(A,C) + \nu_2 (F,B)+\mu_1(D) ~, 
            \\
            I&=\rmd D + \mu_2 (A,D) + \nu_2(F,C) ~,
        \end{aligned}
    \end{equation}
    which read in components as 
    \begin{equation}\label{eq:higher_curvatures_undeformed_cmpnts}
        \begin{aligned}
            \begin{pmatrix}
                G_q \\
                G_u
            \end{pmatrix} &=
            \begin{pmatrix*}[l]
                \rmd C_q \\
                \rmd C_u + C_u \wedge ([A_t,-]) + 2 B_s \wedge (F_t,-) + D_v
            \end{pmatrix*}~,
            \\
            I_v &= \rmd D_v - D_v \wedge ([A_t,-]) - C_u \wedge ([F_t,-])~.
        \end{aligned} 	
    \end{equation}
    The resulting Bianchi identities are
    \begin{equation}
        \begin{aligned}\label{eq:bianchi}
            \rmd G + \mu_2(A,G) -\nu_2 (F, H) - \mu_1(I)=0~, 
            \\
            \rmd I+\mu_2(A,I) - \nu_2(F,G)=0~.
        \end{aligned}
    \end{equation}
    As commonly done in the supergravity literature, we introduce covariantized transformations 
    \begin{equation}
        \begin{aligned}
            \Delta B &= \delta B + \nu_2(\delta A, A) ~,
            \\
            \Delta C &= \delta C + \nu_2 (\delta A, B) ~, 
            \\
            \Delta D &= \delta D + \nu_2(\delta A, C) ~,
        \end{aligned}
    \end{equation}
    and in terms of these, the curvatures transform as 
    \begin{equation}
        \begin{aligned}
            \delta F &= \rmd \delta A + \mu_2 (A,\delta A) + \mu_1(\Delta B) ~,
            \\
            \delta H &= \rmd \Delta B + 2 \nu_2(F,\delta A) - \mu_1(\Delta C)~,
            \\
            \delta G &= \rmd \Delta C + \mu_2 (A,\Delta C) + \nu_2(F,\Delta B) + \nu_2(\delta A, H) + \mu_1(\Delta D)~,
            \\
            \delta I &= \rmd \Delta D + \mu_2 (A,\Delta D) + \nu_2(F,\Delta C) + \nu_2(\delta A, G)~.
        \end{aligned}
    \end{equation}

    \subsection{Field theory Lagrangian}
    
    Having discussed the gauge structure, we can now present the full field theory, as given in~\cite{Saemann:2017zpd}. The field content is mostly arranged into supermultiplets of six-dimensional Minkow\-ski space as indicated in table~\ref{fig:supermultiplets}.
    \begin{table}
        \begin{center}
            \scalebox{0.97}{
                \begin{tabular}{lclcll}
                    \toprule
                    multiplet & field $\Phi$ & field type & values in & SUSY transformation $\delta_{{\rm SUSY},0}$
                    \\
                    \midrule
                    vector & $A$ & 1-form & $\frg_t\oplus \IR^*_p$ & $-\bar \eps \gamma_{(1)} \lambda$
                    \\
                    & $\lambda^i$ & MW spinors & $\frg_t\oplus \IR^*_p$ & $\tfrac{1}{4}\fsl{F}\eps^i - \tfrac12 Y^{ij} \eps_j + \tfrac14 \mu_1 (\phi) \eps^i$
                    \\
                    & $Y^{(ij)}$ & aux.~scalars & $\frg_t\oplus \IR^*_p$ & $-\bar \eps^{(i}\fsl{\nabla} \lambda^{j)}+2\mu_1(\bar\eps^{(i}\chi^{j)})$
                    \\
                    \midrule
                    tensor & $B$ & 2-form & $\IR_r\oplus \IR^*_s$ & $-\bar \eps \gamma_{(2)}\chi-\nu_2( \delta_{\rm SUSY,0} A, A)$
                    \\
                    & $\chi^i$ & MW spinors & $\IR_r\oplus \IR^*_s$ & $\tfrac{1}{8}\fsl{H}\eps^i+\tfrac{1}{4}\dpars\phi\eps^i-\tfrac12 \nu_2(\gamma^\mu\lambda^i,\bar \eps \gamma_\mu \lambda)$
                    \\
                    & $\phi$ & scalar field & $\IR_r\oplus \IR^*_s$ & $\bar \eps \chi$
                    \\
                    \midrule
                    hyper & $q^{i}$ & scalar fields & $\IR^{2N^2}$ & $\epsb^i\psi$
                    \\
                    & $\psi$ & sM spinors & $\IR^{2N^2}$ & $\tfrac12 \nablas q^{i}\eps_i$
                    \\
                    \midrule
                    none & $C$ & 3-form & $\frg^*_u\oplus\IR_q$ & $\nu_2(\bar \eps \gamma_{(3)}\lambda,\phi)-\nu_2(\delta_{\rm SUSY,0} A,B)$
                    \\
                    & $D$ & 4-form & $\frg^*_v$ & $-\nu_2(\delta_{\rm SUSY,0} A,C)$ \\
                    \bottomrule
                \end{tabular}}
        \end{center}
        \caption{The field content with its behavior under Lorentz and supersymmetry transformation as well as its gauge labels. Our convention for the Feynman slash notation is found in appendix~\ref{app:useful} and $\nabla\coloneqq\rmd+[A,-]$. The spinor chiralities are given by $\gamma_7 \epsilon = \epsilon$, $\gamma_7 \lambda = \lambda$, and $\gamma_7 \chi = -\chi$.}
        \label{fig:supermultiplets}
    \end{table}
    
    The $2\times 2N^2$ scalar fields $q^i$ and the $2 N^2$ symplectic Majorana spinors $\psi$ in the hypermultiplet take values in a representation of $\frg_t$ obtained by embedding this Lie algebra into $\frsp(N^2)$, cf.~\cite{Samtleben:2012fb,Saemann:2017zpd}. On the underlying representation space, we have the bilinear pairing
    \begin{equation}\label{eq:symp_pairing}
        \langlec -,-\ranglec: \IR^{2N^2}\times \IR^{2N^2}\rightarrow \IR~,~~~\langlec x,y\ranglec\coloneqq\tfrac12\Omega_{ab}x^a y^b~,
    \end{equation}
    where $\Omega$ is the invariant symplectic form of $\sfSp(N^2)$. In fact, the hypermultiplet will not play any role in our discussion and we shall merely indicate where it would appear in our formulas.
    
    It is also useful to introduce the following supersymmetrically covariantized anti-dual curvatures\footnote{Note that it is normal in higher gauge theory
    	to (linearly) combine differential forms of different degrees into
    	curvature forms.}, whose vanishing implements the expected duality in six dimensions:
    \begin{equation}\label{eq:cov_curvatures}
        \begin{aligned}
            \scH&\coloneqq\tfrac12 (H-\star H)+\nu_2(\bar \lambda,\gamma_{(3)}\lambda)~&&\in \Omega^3(M)\otimes(\IR_r\oplus \IR^*_s)~,
            \\
            \scG&\coloneqq G-\star \nu_2(F,\phi)+\star2\nu_2(\bar \lambda,\gamma_{(2)}\chi)~&&\in\Omega^4(M)\otimes(\frg^*_u\oplus\IR_q)~,
            \\
            \scI&\coloneqq I-\star 4\nu_3(\bar\lambda,\gamma_{(1)}\lambda, \phi)+\scI_{\rm hyper}~&&\in \Omega^5(M)\otimes\frg^*_v~.
        \end{aligned}
    \end{equation}
    Note that $\scH$ is antiself-dual, $\scH=-\star \scH$, and $\scG$ is a duality combination of the 4-form curvature~$G$ and the 2-form curvature~$F$, supersymmetrically completed.

    The Lagrangian composed in~\cite{Saemann:2017zpd} consists of four parts:
    \begin{equation}\label{eq:original_Lagrangian}
        \caL^{\caN=(1,0)}_{\rm PST}=\underbrace{\caL_{\rm gauge}+\caL_{\rm top}+\caL_{\rm hyper}}_{\caL^{\caN=(1,0)}}+\caL_{\rm PST}~.
    \end{equation}
    The first three contain the gauge interactions, a topological part and the matter field interactions. They read as
    \begin{equation}
        \begin{aligned}
            \caL_{\rm gauge} &= - \langle\rmd \phi,\star\rmd \phi\rangle -\star4\langle\bar\chi,\dpars\chi\rangle-\tfrac12\langle H,\star H\rangle-\langle H,\nu_2(\bar\lambda,\star\gamma_{(3)}\lambda)\rangle
            \\
            &\phantom{{}={}}+\big\langle\phi\,,\,\nu_2(F,\star F)-\star2\nu_2(Y_{ij},Y^{ij})+\star4\nu_2(\bar\lambda,\nablas \lambda)\big\rangle
            \\
            &\phantom{{}={}}+4\big\langle \chib,\nu_2(\fsl{F},\lambda)\big\rangle-\star8\big\langle \bar\chi^j,\nu_2(Y_{ij},\lambda^i)\big\rangle~,
            \\
            \caL_{\rm top} &= -\tfrac12\langle\mu_1(C),H\rangle-\tfrac12\big\langle B,\nu_2(F,F)\big\rangle~,\\
            \caL_{\rm hyper} &= -\langlec \nabla q,\star\nabla q\ranglec + \star2\langlec\bar\psi,\nablas\psi\ranglec +\star8\langlec\bar\psi, \lambda_i q^i\ranglec + \star2\,\langlec q^i,Y_{ij} q^j\ranglec~,
        \end{aligned}
    \end{equation}
    where $\langle-,-\rangle$ refers to the natural pairings~\eqref{eq:pairings}, while $\langlec -,-\ranglec$ denotes the pairing~\eqref{eq:symp_pairing}. We also use Feynman slash notation, see appendix~\ref{app:useful}. The covariant derivatives are defined in the obvious way, and we use the same spinor conventions as~\cite{Samtleben:2011fj,Saemann:2017zpd}, see also appendices~\ref{app:susy} and~\ref{app:useful}.
    
    The fourth term $\caL_{\rm PST}$ incorporates a generalization of the PST mechanism developed in~\cite{Saemann:2017zpd}, which ensures that the supersymmetrically covariantized anti-dual curvatures vanish:
    \begin{equation}\label{eq:eom_sd}
        \scH=0~,~~~\scG=0~,~~~\scI=0~.
    \end{equation}
    
    The remaining equations of motions read as 
    \begin{equation}\label{eq:shorthand_eom}
        \begin{aligned}
            \caE_{\lambda_i}&\coloneqq  \nu_2(\nablas\lambda_i,\phi)+\tfrac12\nu_2(\dpars\phi,\lambda_i) -\tfrac12\nu_2(\slasha{F},\chi_i) -\tfrac14\nu_2(\slasha{H},\lambda_i) +\nu_2(Y_{ij},\chi^j)+\caE_{\lambda_i,{\rm hyper}}=0~,\\
            \caE_{Y^{ij}}&\coloneqq\tfrac12\nu_2(Y^{ij},\phi)-\nu_2(\bar\lambda^{(i},\chi^{j)})+\caE_{Y^{ij},{\rm hyper}}=0 ~,\\
            \caE_{\chi_i}&\coloneqq\dpars \chi_i -\nu_2(\slasha{F},\lambda_i)+2\nu_2(Y_{ij},\lambda^j)=0~,\\
            \caE_{\phi}&\coloneqq\square \phi  -\star\nu_2(F,\star F) - 2\nu_2(Y_{ij},Y^{ij})+4\nu_2(\bar\lambda,\nablas\lambda)=0~,
        \end{aligned}
    \end{equation}
    where $\square\coloneqq\dpar^\mu\dpar_\mu$, together with again evident expressions for the equations of motion of the fields in the hypermultiplet.
    
    \subsection{Drawbacks of the PST mechanism}\label{ssec:problems_PST}
    
    The generalization of the PST mechanism incorporated into the field theory by $\caL_{\rm PST}$ comes with a number of disadvantages. Explicitly, this part of the Lagrangian reads as 
    \begin{equation}\label{eq:Lagrangian_PST}
        \caL_{\rm PST} = \tfrac12\big\langle\iota_{V}\scH,\scH\big\rangle\wedge v+\langle \Phi(\iota_{V} \star \scG), \star~\iota_{V}\star\scG\rangle~,
    \end{equation}
    where $v$ is a nowhere vanishing exact auxiliary 1-form, and $V$ its corresponding dual vector field:
    \begin{equation}
        v = v_\mu \rmd x^\mu = \rmd a,~~~\iota_V v=1~,~~~\iota_V \star v=0~.
    \end{equation}
    Furthermore, $\Phi$ is the map 
    \begin{equation}
        \Phi:\frg_u^*\oplus \IR_q\rightarrow \frg_t\oplus \IR^*_p,~~~\Phi(u+q)\coloneqq\frac{1}{\phi_s}(u,-)~,
    \end{equation}
    where $(-,-):\frg^*\times \frg^*\rightarrow \IR$ is the inverse of the metric $(-,-)$ on $\frg$ and $\phi_s\coloneqq\phi|_{\IR^*_s}$. 
    
    An obvious problem is that $\Phi$ is not defined whenever $\phi_s$ vanishes. This seems to be a common feature; see also the paper~\cite{Bandos:2013jva} where the PST mechanism was incorporated in a similar model. This problem subsides by treating the limit $\phi_s$ carefully. This limit, however, is particularly interesting as it represents the point at which the $(1,0)$-model would be expected to gain additional symmetry and to turn into the $(2,0)$-theory. The form of the Lagrangian is therefore at least inconvenient.
    
    Perhaps the main disadvantage of implementing self-duality and, more generally, the equations~\eqref{eq:eom_sd} using the PST mechanism is the additional PST field, which appears in a non-standard manner and which thus requires a non-standard quantization procedure. Given the simplicity of Sen's action~\eqref{eq:sen_action}, it is therefore reasonable to try to replace $\caL_{\rm PST}$ with an adaptation of this mechanism. Moreover, the Lagrangian with which we end up is a simple extension of the Lagrangian $\caL^{\caN=(1,0)}$ by fields which are on-shell physically trivial and this is clearly simpler and more convenient than the PST mechanism.
    
    \section{Self-duality from Lagrange multipliers}
    
    This section contains our main result: a fully supersymmetric action whose equations of motion include the appropriate duality relations $\scH=\scG=\scI=0$ for the curvature forms.
    
    \subsection{Duality equations from an action principle}
    
    Our starting point is the Lagrangian $\caL_0$ for the non-gauge fields in the tensor multiplet as well as the fields in the vector (and hyper) multiplet,
    \begin{equation}
        \begin{aligned}
            \caL_0&=- \langle\rmd \phi,\star\rmd \phi\rangle -\star4 \langle\bar\chi,\dpars\chi\rangle+\star4\big\langle \chib,\nu_2(\fsl{F},\lambda)\big\rangle-\star8\big\langle \bar\chi^j,\nu_2(Y_{ij},\lambda^i)\big\rangle
            \\
            &\hspace{1cm}+\big\langle\phi\,,\,\nu_2(F,\star F)-\star2\nu_2(Y_{ij},Y^{ij})+\star4\nu_2(\bar\lambda,\nablas \lambda)\big\rangle + \ldots~.
        \end{aligned}
    \end{equation}
    In order to implement an analogue of Sen's mechanism, we merely introduce a self-dual 3-form 
    \begin{equation}
        \beth_s\in \Omega^3_+(M)\otimes \IR^*_s
    \end{equation}
    and consider the Lagrangian 
    \begin{equation}
        \begin{aligned}
            \caL_\beth&= -H_s \wedge \star \scH_r + H_s \wedge C_q - \beth_s \wedge \scH_r 
            \\
            &= ~(\rmd B_s)^{+} \wedge \left(\rmd B_r + {\rm cs}(A)_r + (\bar{\lambda}_t, \gamma_{(3)} \lambda_t)\right) + (\rmd B_s)^- \wedge C_q^{+}   
            \\
            &\hspace{1cm}-\beth_s \wedge (\rmd B_r + {\rm cs}(A)_r - C_q + (\bar{\lambda}_t, \gamma_{(3)} \lambda_t))~,
        \end{aligned}
    \end{equation}
    where $C_q^\pm$ and $(\rmd B_s)^\pm$ denote self-dual and antiself-dual parts of the 3-forms $C_q$ and $\rmd B_s$ and
    \begin{equation}
        {\rm cs}(A)_r\coloneqq(A_t,\rmd A_t)+\tfrac13(A_t,[A_t,A_t])~.
    \end{equation}
    Varying the action with respect to $\beth_s$ leads to the equation of motion $\scH_r=0$. Furthermore, varying with respect to $C_q$ leads to the equation
    \begin{equation}
        0=\beth_s+H_s^-~~~\Rightarrow~~~\beth_s=0~,~~~H_s^-=\scH_s=0~.
    \end{equation}
    We see that $\beth_s$ vanishes by the equations of motion, which is different from the actions discussed in section~\ref{sec:self-duality_and_homotopy_algebras}, where $\beth$ either decoupled or was gauge trivial. Here, $\beth_s$ simply does not contain any on-shell degrees of freedom and no further arguments are required. 
    
    In order to enforce the equation $\scG=0$, we introduce a further field
    \begin{equation}
        \daleth_t\in \Omega^2(M)\otimes \frg_t
    \end{equation}
    and include the Lagrangian 
    \begin{equation}\label{eq:daleth-terms}
        \begin{aligned}
            \caL_\daleth & = ~\scG_u(\daleth_t) -B_s \wedge (\daleth_t, \daleth_t) +\phi_s (\daleth_t, \star \daleth_t) 
            \\
            &=~(\rmd C_u)(\daleth_t) + C_u\wedge([A,\daleth_t]) +  2 B_s \wedge (F_t, \daleth_t) + D_v(\daleth_t) - 2 \phi_s (\daleth_t , \star F_t) 
            \\
            &\hspace{1cm} -\star4\chib_s(\fsl{\daleth}_t,\lambda_t)-B_s \wedge (\daleth_t, \daleth_t) +\phi_s (\daleth_t, \star \daleth_t)~.
        \end{aligned}
    \end{equation}
    The motivation for considering the term $\scG_u (\daleth_t)$ should be clear; the remaining two terms are derived when trying to close supersymmetry as we shall argue later. For now, we note that varying the action with respect to $D_v$ yields the equation of motion $\daleth_t=0$; similarly, varying with respect to $\daleth_t$ leads to the equation of motion 
    \begin{equation}\label{eq:eom_G}
        0=\scG_u -2B_s \wedge (\daleth_t,-) + 2 \phi_s \star (\daleth_t,-)=\scG_u~.
    \end{equation}
    Again, we implemented a desired equation of motion and the new field we added for this turns out not to carry any on-shell degrees of freedom. 
    
    Varying $\caL_{\daleth}+\caL_{\beth}$ by $B_s$, we obtain the equation $\scG_q=\rmd C_q=0$. The only remaining equation $\scI_v=0$ is obtained by varying the total action with respect to $A_t$ and comparing the resulting terms with those obtained from 
    \begin{equation}
        \rmd \scG_u+\scG_u\wedge ([A,-])=I_v+\Big(2H_s\wedge F_t-2\nabla \left(\phi_s \star F_t\right) + 4\nabla \left(\lambdab_t \star \gamma_{(2)} \chi_s \right)~,~-\Big)=0~,
    \end{equation}
    which is trivially true due to~\eqref{eq:eom_G} and the Bianchi identity~\eqref{eq:bianchi} for $ G_u $.
    
    The total action of our field theory reads as 
    \begin{equation}\label{eq:new_Lagrangian}
        S^{\caN=(1,0)}_{\rm TF}=\int_{M} \rmd^6x~\caL^{\caN=(1,0)}_{\rm TF}\ewith \caL^{\caN=(1,0)}_{\rm TF} = \caL_{\beth} + \caL_{\daleth} + \caL_{0}~,
    \end{equation}
    and varying with respect to the various fields, we obtain the equations of motion listed in table~\ref{fig:eom}. These are indeed all the expected equations.
    
    We also note that upon putting $\beth_s=0$ and $\daleth_t=0$, we obtain the original Lagrangian $\caL^{\caN=(1,0)}$ given in~\eqref{eq:original_Lagrangian}. This is rather clear for the matter field couplings. As a short computation shows, the topological term $\caL_{\rm top}$ indeed combines with $\caL_{\rm gauge}$ to $\caL_0+\caL_\beth|_{\beth_s=0}$.
        
    \begin{table}
        \begin{center}
            \scalebox{0.95}{\begin{tabular}{clr}
                \toprule
                $\Phi$ & $\frac{\delta S}{\delta \Phi}$ & simplified eom
                \\
                \midrule
                $\beth_s$ & $\scH_r$ & $\scH_r=0$
                \\
                $\daleth_t$ & $\scG_u -2B_s \wedge (\daleth_t,-) + 2 \phi_s \star (\daleth_t,-)$ & $\scG_u=0$
                \\
                \midrule
                $C_q$ & $H^{-}_s+\beth_s$ & $\begin{cases}
                    \scH_s=0 \\
                    \,\,\beth_s=0
                \end{cases}$\hspace*{-0.43cm}
                \\
                $D_v$ & $\daleth_t$ & $\daleth_t = 0 $
                \\
                \midrule
                $B_s$ & $-\rmd (\scH_{r}  + C_q) + 2(F_t, \daleth_t)-(\daleth_t,\daleth_t)$ & $\scG_q= 0$
                \\
                $B_r$ & $\rmd \left( H^{+}_s-\beth_s \right) = 0$ &$ 0=0 $  
                \\
                $C_u$ & $\nabla \daleth_t$ &  $ 0=0 $ 
                \\
                $\phi_s$ & $\square \phi_s -\star (F_t,\star F_t) -2(Y_{tij},Y^{ij}_t)+4(\lambdab_t, \nablas \lambda_t)$ & $\caE_{\phi_r}=0$
                \\
                $\phi_r$ & $\square \phi_r$ & $\caE_{\phi_s}=0$
                \\
                $\chi_r $ & $\dpars \chi_s$ & $\caE_{\chi_r}=0$
                \\
                $\chi_s$ & $ \dpars \chi_{ri} - (\fsl{F}_{\!t}-\slasha{\daleth_t}, \lambda_{ti}) + 2 (Y_{tij}, \lambda^j_t)$ & $\caE_{\chi_s}=0$
                \\
                \midrule
                \\
                $A_t$ & $-\nabla \left(\phi_s \star F_t\right) + 2\nabla \left(\lambdab_t \star \gamma_{(2)} \chi_s \right) + F_t\wedge H_s +2\phi_s [\lambdab, \star \gamma_{(1)} \lambda]$ & $\scI_v=0$
                \\ 
                & $+ \text{ terms containing } \daleth+  \text{ terms containing } \beth$ +\ldots
                \\
                $A_s$ & $0$ & 
                \\
                $\lambda_t$ & $2\phi_s \nablas \lambda_{ti} + (\dpars \phi_s) \lambda_{ti} - \tfrac12 \big(\slasha{H}_{\!s} - \fsl{\beth}_s\big) \lambda_{ti} - \big(\fsl{F}_{\!t}-\fsl{\daleth_t}\big) \chi_{si} + 2 Y_{ij} \chi_s^{j}+\ldots$ & $\caE_{\lambda_t}=0$
                \\
                $\lambda_s$ & $0$ & 
                \\
                $Y_{tij}$ & $\phi_s Y_t^{ij} - 2 \lambdab_t^{(i} \chi_s^{j)}+\ldots$ & $\caE_{Y^{ij}_t}=0$
                \\
                \bottomrule
            \end{tabular}}
        \end{center}
        \caption{The (normalized) variations of $S$ with respect to the various fields and the resulting simplified equations of motion. The expressions $\caE_\chi$, $\caE_\phi$, $\caE_\lambda$, and $\caE_{Y^{ij}}$ are defined in~\eqref{eq:shorthand_eom}. Ellipses denote terms originating from the hypermultiplets. We use again Feynman slash notation, cf.~appendix~\ref{app:useful}. Note that the equation $\scI_v=0$ is obtained from the equation for $\Phi=A_t$, combined with the equation for $\Phi=\daleth_t$.}
        \label{fig:eom}
    \end{table}
    
    \subsection{Supersymmetry}
    
    Let us discuss the supersymmetry of the Lagrangian~\eqref{eq:new_Lagrangian} in some more detail. For this, we decompose it into a part $\caL^{\caN=(1,0)}$ independent of $\beth_s$ and $\daleth_t$ as well as the remainder $\caL_{\rm TF}$:
    \begin{equation}
        \caL^{\caN=(1,0)}_{\rm TF}=\caL^{\caN=(1,0)}+\caL_{\rm TF}~.
    \end{equation}
    We know that the Lagrangian $\caL^{\caN=(1,0)}$ is supersymmetric under the supersymmetry transformations $\delta_{{\rm SUSY},0}$ listed in table~\ref{fig:supermultiplets}, because it is essentially a specialization of the model of~\cite{Samtleben:2011fj}.
    
    The Lagrangian $\caL_{\rm TF}$ contains terms which transform non-trivially under $\delta_{{\rm SUSY},0}$ and to ensure supersymmetry of $\caL^{\caN=(1,0)}_{\rm TF}$ we need to shift the transformations of certain fields. We have 
    \begin{equation}
        \caL_{\rm TF} = -\beth_s \wedge \scH_r +  \scG_u(\daleth_t) -B_s \wedge (\daleth_t, \daleth_t) +\phi_s (\daleth_t, \star \daleth_t) ~,
    \end{equation}
    and we note that the arising additional terms are very similar to those in $\caL^{\caN=(1,0)}$ containing gauge curvatures. In the cancellation of supersymmetry transformations, the latter arise from the supersymmetry transformations of the spinor fields, and it is therefore natural to try to shift their transformations in order to cancel the terms in $\caL_{\rm TF}$.
    
    Introducing 
    \begin{subequations}\label{eq:shifted_SUSY}
        \begin{equation}
            \delta_{\rm SUSY}=\delta_{{\rm SUSY},0}+\delta_{{\rm SUSY},1}~,
        \end{equation}
        we find that the necessary shifts are 
        \begin{equation}
            \begin{aligned}
                \delta_{{\rm SUSY},1}\,\chi_s &= -\tfrac18 \slasha{\beth}_s \eps~,
                \\
                \delta_{{\rm SUSY},1}\,\lambda_t &= -\tfrac14 \slasha{\daleth_t} \eps~,
                \\
                \delta_{{\rm SUSY},1}\,C& =2\hat\nu_2(\delta A, \daleth)~,
            \end{aligned}
        \end{equation}
        and the auxiliary fields are supersymmetry singlets,
        \begin{equation}
            \delta_{\rm SUSY}\beth_s= 0~,~~~\delta_{\rm SUSY}\daleth_t=0~.
        \end{equation}
    \end{subequations}
    
    Because the auxiliary fields vanish on-shell, $\beth_s=0$ and $\daleth_t=0$, it is immediately obvious that this shifted supersymmetry algebra closes on-shell if the original supersymmetry algebra does, which is the case.
    
    Verification of supersymmetry is then a tedious but in principle straightforward exercise. We collect a number of useful relations and equations we used in verifying the supersymmetry of the action corresponding to  $\caL^{\caN=(1,0)}_{\rm TF}$ in appendix~\ref{app:useful}.
    
    Finally, we note that the BPS states of this new action and the new supersymmetry transformation still comprise the non-abelian self-dual strings introduced in~\cite{Saemann:2017rjm}.
    
    \section*{Acknowledgements}
    
    We would like to thank Leron Borsten, Hyungrok Kim, and Martin Wolf for interesting questions. The work of CS was partially supported by the Leverhulme Research Project Grant RPG-2018-329 ``The Mathematics of M5-Branes.''
    
    \appendices
    
    \subsection{Gauge structure dictionary to supergravity literature}\label{app:dictionary}
    
    Our formulas can directly be compared to the supergravity literature and in particular to the papers~\cite{Samtleben:2011fj,Samtleben:2012fb,Bandos:2013jva}. The structure constants in these papers are related to our $EL_\infty$-algebra $\aghsk^\omega$ as indicated in table~\ref{fig:dictionary}.
    \begin{table}
        \begin{center}
            \begin{tabular}{@{}lll@{}}
                \toprule
                & Notation~\cite{Samtleben:2011fj,Bandos:2013jva} & Translated to $\aghsk^\omega$\\
                \midrule
                Indices & $T^r$ & $T^\alpha+T_q\in \frg_t\oplus \IR^*_p$\\
                ($T$: general obj.)& $T^I$ & $T_r+T_s \in \IR_r\oplus \IR^*_s$\\
                & $T_r$ & $T_\alpha+T_p\in \frg^*_u\oplus \IR_q$\\
                & $T_\alpha$ & $T_\alpha\in \frg^*_v$\\
                \midrule
                Structure const.\ & $h^r_I$ & $\mu_1=\sfid:\IR^*_s\rightarrow \IR^*_p$\\
                & $-g^{Ir}$ & $\mu_1=\sfid:\IR_q\rightarrow \IR_t$\\
                & $k_r^\alpha$ & $\mu_1=\sfid:\frg^*_v\rightarrow \frg^*_u$\\
                & $-f_{st}{}^r$ & $\mu_2:\frg_t\wedge \frg_t\rightarrow \frg_t:~ \mu_2(\xi_1,\xi_2)\coloneqq [\xi_1,\xi_2]$\\
                & $d_{rs}^I$ & $\nu_2:\frg_t\otimes \frg_t \rightarrow \IR_r: (\xi_1,\xi_2)$\\
                & $-b_{Irs}$ & $\nu_2:\frg_t\otimes \IR^*_s\rightarrow \frg^*_u: \nu_2(\xi,s)\coloneqq 2\langle(-,\xi),s\rangle$\\
                & $c^t_{\alpha s}$ & $\nu_2:\frg_r\otimes \frg^*_u\rightarrow \frg^*_v: \nu_2(\xi,\zeta)\coloneqq \zeta([-,\xi])$\\
                \bottomrule
            \end{tabular}
        \end{center}
        \caption{The relation between the structure constants employed in~\cite{Samtleben:2011fj,Samtleben:2012fb,Bandos:2013jva} and our $EL_\infty$-algebra maps.}
        \label{fig:dictionary}
    \end{table}
    
    Moreover, the metric in~\cite{Samtleben:2011fj} on the tensor multiplet fields is 
    \begin{equation}
        (\eta_{IJ})=\begin{pmatrix}
            0 & -1 \\
            -1 & 0 
        \end{pmatrix}~.
    \end{equation}
    Finally, the structure constants $k_r^\alpha$ are not uniquely fixed in this framework, and we choose the simplest possibility. We used slightly different conventions to those in~\cite{Saemann:2017zpd}.
    
    \subsection{Supersymmetry in six dimensions}\label{app:susy}
    
    We work on six-dimensional Minkowski space $M=\IR^{1,5}$ with a ``mostly plus'' Minkowski metric,
    \begin{equation}
        (\eta_{\mu\nu})={\rm diag}(-1,+1,+1,+1,+1,+1)~,
    \end{equation}
    and the gamma matrices $(\gamma_\mu)$ form a representation of the Clifford algebra
    \begin{equation}
        \{\gamma_{\mu},\gamma_{\nu}\}=2\eta_{\mu \nu}~.
    \end{equation}
    For the Levi--Civita symbol, we use the convention
    \begin{equation}
        \eps_{012345}=1~,
    \end{equation}
    and we always work with normalized antisymmetrization brackets $[\mu_1\ldots \mu_n]$ involving the usual factors of $\frac{1}{n!}$. For example,
    \begin{equation}
        \gamma_{\mu \nu}\coloneqq\gamma_{[\mu\nu]}\coloneqq\tfrac{1}{2}\left(\gamma_{\mu}\gamma_{\nu}-\gamma_{\nu}\gamma_{\mu}\right)~.
    \end{equation}
    
    In six dimension, we have the following dualities between complementary ranks
    \begin{equation}
        \gamma_{\mu_1\ldots\mu_r}\gamma_7=\frac{t_r}{(6-r)!} \gamma^{\nu_1\ldots\nu_{6-r}} \eps_{\nu_1\ldots\nu_{6-r}\mu_1\ldots\mu_r}~,
    \end{equation}
    where
    \begin{equation}
        t_r=-(-1)^{\frac{r(r+1)}{2}} = 
        \begin{cases}
            -1 ~: \quad r=0,3,4~, \\
            +1 ~: \quad r=1,2,5,6~,
        \end{cases}
    \end{equation}
    cf.~e.g.~\cite{Freedman:2012zz} for more details. Defining 
    \begin{equation}
        \gamma_{(r)}\coloneqq\frac{1}{r!}\gamma_{\mu_1\ldots\mu_r}dx^{\mu_1}\wedge\ldots\wedge dx^{\mu_r}~,
    \end{equation}
    we can write this identity as 
    \begin{equation}
        t_r\gamma_{(r)}\gamma_7=\star \gamma_{(6-r)} ~.
    \end{equation}
    We define Majorana conjugates of spinors in a standard way, $ \bar{\lambda}=\lambda^T C $ where the charge conjugation matrix $ C $ is symmetric. This matrix is important for symmetry properties of gamma matrices
    \begin{equation}
        (C\gamma^{[M_r]})^T=-t_rC\gamma^{[M_r]}~,
    \end{equation}
    where $M_r$ is a multiindex containing $r$ simple indices, $\gamma^{[M_r]}=\gamma^{\mu_1\ldots\mu_r} $.
    For Gra{\ss}mann-valued spinors we have the symmetries
    \begin{equation}
        \bar{\lambda}_1^i \gamma^{[M_{r_1}]}\ldots\gamma^{[M_{r_p}]}\lambda_2^j=t_0^{p-1}t_{r_1}\ldots t_{r_p} \bar{\lambda}_2^j \gamma^{[M_{r_p}]}\ldots\gamma^{[M_{r_1}]} \lambda_1^i~,
    \end{equation}
    and a useful special case is
    \begin{equation}
        \bar{\lambda}_1^i \gamma^{\mu_1\ldots\mu_r}\lambda_2^j=t_r \bar{\lambda}_2^j \gamma^{\mu_1\ldots\mu_r}\lambda_1^i~.
    \end{equation}
    
    In order to resolve products of gamma matrices such as $\gamma^{\mu \nu \rho} \gamma_{\alpha \beta}$, we first write down the completely antisymmetrized product of gamma matrices and then contract neighboring upper and lower indices into Kronecker deltas multiplied by an appropriate combinatorial factor. The upper and lower indices in resulting terms need to be antisymmetrized. There are no extra signs arising if indices are contracted appropriately. For example,
    \begin{equation}
        \gamma^{\mu \nu \rho} \gamma_{\alpha \beta}= \gamma^{\mu \nu \rho}_{\phantom{\mu \nu \rho} \alpha \beta} + 6 \gamma^{[\mu \nu}_{\phantom{[\mu \nu} [\beta} \delta^{\rho]}_{\alpha]} + 6 \gamma^{[\mu} \delta^{\nu}_{[\beta} \delta^{\rho]}_{\alpha]} ~.
    \end{equation}
    The factor of~6 in the second term is obtained from~$3=\frac{3!}{2}$, coming from the upper antisymmetrization bracket taking into account the already antisymmetrized indices $ \mu\nu $, times~$2!$ from the lower bracket. Another useful example is
    \begin{equation} \label{gp13}
        \gamma_{\mu \nu \rho} \gamma_{\sigma} = \gamma_{\mu \nu \rho \sigma} + 3\gamma_{[\mu \nu} \eta_{\rho] \sigma} \quad \Longrightarrow \quad \begin{cases}
            \{\gamma_{\mu \nu \rho}, \gamma_{\sigma}\}=6\gamma_{[\mu \nu} \eta_{\rho] \sigma}~, \\
            \left[ \gamma_{\mu \nu \rho}, \gamma_{\sigma} \right]=2\gamma_{\mu \nu \rho \sigma}~.
        \end{cases}
    \end{equation}
    
    The R-symmetry group is $\sfSp(1)\cong \sfSU(2)$ and spinors, e.g.~$\lambda$, carry corresponding indices, e.g.~$\lambda^i$, $i\in \{1,2\}$. Indices are always contracted in the evident way with the convention, e.g.
    \begin{equation}
        (\bar\lambda\ldots\lambda)=\bar\lambda^i \ldots \lambda_i\ewith \lambda^i=\eps^{ij}\lambda_j~,
    \end{equation}
    and we use $\eps^{12}=\eps_{12}=1$.

    \subsection{Useful identities}\label{app:useful}
    
    Below we collect a number of helpful identities and results that we used in verifying the supersymmetry of the action~$S^{\caN=(1,0)}_{\rm TF}$. Our convention for the spinor chiralities are those of~\cite{Samtleben:2011fj}. That is,
    \begin{equation}
        \gamma_7 \epsilon = \epsilon ~, \quad \gamma_7 \lambda = \lambda ~, \quad \gamma_7 \chi = -\chi ~.
    \end{equation}
    A useful Fierz identity is
    \begin{equation}
    \eps^j_2\epsb^i_1=\frac{1}{4}\left(\epsb_1^i\gamma^{\mu}\eps_2^j\gamma_{\mu} -\tfrac{1}{12}\epsb_1^i\gamma^{\mu \nu \rho} \eps_2^j \gamma_{\mu \nu \rho}\right) \frac{1-\gamma_7}{2} ~.
    \end{equation}
    With the gamma matrix gymnastics reviewed in the previous appendix, we obtain the identities
    \begin{equation}
        \gamma_{(2)} \wedge \rmd \chi = \tfrac{1}{2} \gamma_{(3)} \dpars \chi + \tfrac{1}{2} \dpars \gamma_{(3)} \chi
        ~,~~~ 
        \star \left(\gamma_{(2)} \wedge \rmd \chi \right) = -\tfrac{1}{2} \gamma_{(3)} \dpars \chi + \tfrac{1}{2} \dpars \gamma_{(3)} \chi ~.
    \end{equation}
    We use Feynman slash notation which is defined with the evident normalization, e.g.
    \begin{equation}
        \fsl{F}=\tfrac12 \gamma_{\mu}\gamma_{\nu} F^{\mu\nu}=\tfrac12 \gamma_{\mu\nu}F^{\mu\nu}\eand \slasha{H}=\tfrac{1}{3!} \gamma_{\mu \nu \rho} H^{\mu \nu \rho}~.
    \end{equation}
    We often decompose a 3-form such as $H$ into its self-dual and antiself-dual  parts $ H= H^++H^-$,
    \begin{equation}
        H^{\pm}=\pm \star H^{\pm} \quad \Longleftrightarrow \quad  H^{\pm}_{\mu \nu \rho}=\pm \tfrac{1}{3!} \eps_{\alpha \beta \gamma \mu \nu \rho} H^{\pm ~\alpha \beta \gamma} ~,
    \end{equation}
    and using this decomposition, we find the identities
    \begin{equation} \label{app:3form-partial}
        \begin{aligned}
            \dpars \slasha{H} &= -\star \left[ \gamma_{(4)} \wedge \star \rmd H + \gamma_{(2)} \wedge \rmd \star H \right]  \\
            &= -\star \left(\rmd H^+ \wedge \gamma_{(2)} \right) (\gamma_7+1) -\star \left(\rmd H^- \wedge \gamma_{(2)} \right) (\gamma_7-1)  ~, \\
            \slasha{H} \dpars &= -\star \left[ H\wedge \rmd \star \gamma_{(4)} + H \wedge \star \left(\gamma_{(2)}\wedge \rmd \right) \right] \\
            &= -\star \left[ H^+\wedge \gamma_{(2)} \wedge \rmd \right] \left( \gamma_7 -1 \right) -\star \left[ H^-\wedge \gamma_{(2)} \wedge \rmd \right] \left( \gamma_7 +1 \right) ~.
        \end{aligned}
    \end{equation}
    For a 2-form $ F $ valued in $ \frg $, we have the identity
    \begin{equation}
    	\star(\fsl{F},\fsl{F})=(F,F)\wedge \gamma_{(2)} \gamma_7 - (F,\star F)~,
    \end{equation}
    where $ (-,-) $ is the metric on $ \frg $.
    
    The supersymmetry transformation of our Lagrangian can then be decomposed as follows
    \begin{equation}
        \delta_{{\rm SUSY}}\caL^{\caN=(1,0)}_{\rm TF}=\delta_{{\rm SUSY},1}\caL^{\caN=(1,0)}+\delta_{{\rm SUSY}}\caL_{\rm TF}~,
    \end{equation}
    where we have used that $\delta_{{\rm SUSY},0}\caL^{\caN=(1,0)}=0 $. The individual terms now transform as follows:
    \begin{equation}
        \begin{aligned}
            \delta_{{\rm SUSY},1}\caL^{\caN=(1,0)} &=~ H_s^+ \wedge \delta_{{\rm SUSY},1} (\lambdab , \gamma_{(3)} \lambda) + H_s^- \wedge \delta_{{\rm SUSY},1}C_q -\star 4\delta_{{\rm SUSY},1} \chib_s \dpars \chi_r 
            \\
            &\hspace{0.5cm}+\star 4 \phi_s \delta_{{\rm SUSY},1}(\lambdab_t, \nablas \lambda_t) + \star 4 (\delta_{{\rm SUSY},1} \chib_s^i) \left[( \fsl{F}_{\!t}, \lambda_{ti}) -2(Y_{tij}, \lambda^{j}_t) \right] \\
            &\hspace{0.5cm}+\star 4\chib_s^i \left[(\fsl{F}_{\!t},\delta_{{\rm SUSY},1} \lambda_{ti})-2(Y_{tij}, \delta_{{\rm SUSY},1} \lambda_{t}^j)\right]~, \\
            \delta_{{\rm SUSY}}\caL_{\rm TF} &= ~-\beth_s \wedge \delta_{{\rm SUSY}} \scH_{r} + (\delta_{{\rm SUSY}} \scG_u)(\daleth_t) -\delta_{{\rm SUSY}} B_s \wedge (\daleth_t, \daleth_t)
            \\
            &\hspace{0.5cm} +\delta_{{\rm SUSY}}\phi_s (\daleth_t, \star \daleth_t)~.
        \end{aligned}
    \end{equation}
    We evaluate and collect individual terms below:
    \begin{equation}
        \begin{gathered}
            -\star 4\delta_{{\rm SUSY},1} \chib_s \dpars \chi_r + \star 4 (\delta_{{\rm SUSY},1} \chib_s^i) \left[( \fsl{F}_{\!t}, \lambda_{ti}) -2(Y_{tij}, \lambda^{j}_t) \right] = \star\tfrac12\epsb\fsl{\beth}_s\caE_{\chi_r}~,
            \\[0.2cm]
            -\beth_s\wedge\delta_{{\rm SUSY}}\scH_{r} = -\star\tfrac12\epsb\fsl{\beth}_s\caE_{\chi_r}-\star\tfrac12\epsb\fsl{\beth}_s(\fsl{\daleth}_t,\lambda_{t}) ~, 
            \\[0.2cm]
            \star 4 \phi_s \delta_{{\rm SUSY},1}(\lambdab_t, \nablas \lambda_t)+\star 4\chib_s^i \left[(\fsl{F}_{\!t},\delta_{{\rm SUSY},1} \lambda_{ti})-2(Y_{tij}, \delta_{{\rm SUSY},1} \lambda_{t}^j)\right] = \hspace{2cm}
            \\
            \hspace{8.5cm}\star \epsb (\fsl{\daleth}_t,\caE_{\lambda_t})+\star\tfrac12\epsb(\fsl{\daleth}_t,\fsl{H}_{\!s}\lambda_{t})~, 
            \\[0.2cm]
            H_s^+ \wedge \delta_{{\rm SUSY},1} (\lambdab , \gamma_{(3)} \lambda) + H_s^- \wedge \delta_{{\rm SUSY},1}C_q =\hspace{5cm}
             \\
             \hspace{4cm}-\star\tfrac12\epsb (\fsl{\daleth}_t,\fsl{H}_{\!s}\lambda_{t}) -2H_s^-\wedge(\epsb \gamma_{(1)} \lambda_{t}, \daleth_t)~,
            \\[0.2cm]
            (\delta_{{\rm SUSY}}\scG_u)(\daleth_t)-\delta_{{\rm SUSY}} B_s \wedge (\daleth_t, \daleth_t)+\delta_{{\rm SUSY}}\phi_s (\daleth_t, \star \daleth_t)=\hspace{3.5cm}
            \\
            \hspace{4.5cm}-\star\epsb  (\fsl{\daleth}_t, \caE_{\lambda_t}) +2 H_{s}^-\wedge (\epsb \gamma_{(1)} \lambda_{t}, \daleth_t)+ \star\tfrac12\epsb \fsl{\beth}_s(\fsl{\daleth}_t,\lambda_t)~.
        \end{gathered} 
    \end{equation}
    It is then clear that 
    \begin{equation}
    	\delta_{{\rm SUSY}}\caL^{\caN=(1,0)}_{\rm TF}=0~.
    \end{equation}

    \bibliography{bigone}
    
    \bibliographystyle{latexeu}
    
\end{document}